\documentclass[preprint,12pt]{elsarticle}
\journal{Annals of Physics}
\usepackage{amsmath,amssymb}
\usepackage{microtype}
\usepackage{graphicx}
\usepackage{physics}
\usepackage{hyperref}
\hypersetup{hidelinks}

\begin{document}
\begin{frontmatter}
\title{Dynamic Eikonal Attenuation and Mutual Coherence of Quantum Particles due to Equilibrium Coulomb Fluctuations}
\author[hse,ras]{Yury A. Budkov\corref{cor1}}
\cortext[cor1]{Corresponding author}
\ead{ybudkov@hse.ru}
\affiliation[hse]{organization={Laboratory of Computational Physics, HSE University},
 addressline={Tallinskaya st. 34}, postcode={123458}, city={Moscow}, country={Russia}}
\affiliation[ras]{organization={Frumkin Institute of Physical Chemistry and Electrochemistry, Russian Academy of Sciences},
 addressline={31-4 Leninsky Prospect}, postcode={119071}, city={Moscow}, country={Russia}}

\begin{abstract}
We develop a dynamic theory of eikonal attenuation and mutual coherence for a quantum particle propagating through equilibrium Coulomb fluctuations. The one-particle length $\ell$ is the attenuation length of the ensemble-averaged retarded propagator, not an Anderson localization length or a transport mean free path. For a collisionless classical one-component plasma (OCP), the potential spectrum follows from the fluctuation--dissipation theorem and the RPA dielectric function. In suitable plasma parameter windows, both fast and slow probe regimes are physically meaningful. Fast probes recover the Coulomb logarithm with the dynamical infrared length $v/\omega_{\rm p}$. Slow probes yield $G(v)\propto v$ and, in the controlled weak branch, $\ell\propto k$. We derive an exact finite-thickness mutual-coherence expression and a closed slow-regime transverse scaling function. Within the controlled slow-OCP and thick-slab window, the logarithmic two-ray visibility scales as $v^{-1}$ at fixed transverse separation, in contrast to the $v^{-2}$ dependence for frozen finite-slab fluctuations. For a symmetric diffusive electrolyte, linearized stochastic density functional theory gives the exact on-shell result $G_{\rm el}(v)=(m^2C/\hbar^4)\operatorname{arsinh}[v/(2D\kappa)]$, where $C=k_{\mathrm B}Tq_0^2/(4\pi\varepsilon_b)$. For ordinary liquid electrolytes, the slow diffusive limit is inaccessible within the spatial eikonal regime and is therefore a formal asymptote. The physically relevant electrolyte limit is instead that of a fast probe, $v\gg D\kappa$, where the diffusive charge relaxation generates an intrinsic infrared length $v/(D\kappa^2)$ absent from the quasi-static theory. The two media have identical Gaussian equal-time potential correlations but different dynamical dephasing laws. Their comparison separates robust fast-probe regularization from mathematically informative, yet not generally observable, slow-electrolyte coherence scaling.
\end{abstract}
\end{frontmatter}

\section{Introduction}

Attenuation of an ensemble-averaged propagator, phase decorrelation, and Anderson localization are distinct manifestations of wave propagation in a random environment. Anderson localization is defined through typical amplitudes, transfer-matrix Lyapunov exponents, or transport observables~\cite{anderson1958absence,kramer1993localization,lee1985disordered}. Scaling approaches can relate localization characteristics to the correlation structure of the random potential~\cite{Satanin1991}, while correlated disorder more generally can produce nonstandard propagation and localization behavior~\cite{izrailev2012anomalous}. Long-range Coulomb disorder and interaction effects near metal--insulator transitions have played a central role in the theory of disordered electronic systems~\cite{shklovskii2013electronic,shklovskii2024half}. By contrast, an averaged one-particle Green function may decay exponentially even when no rigorous localization length has been established. This distinction is particularly important for Coulomb media, where long-wavelength potential fluctuations can make the averaged propagator infrared sensitive.

Coulomb fluctuations occur in plasmas, electrolyte solutions, ionic liquids, ultracold plasma systems, and Coulomb glasses~\cite{sous2019many,vojta1998quantum}. In a classical one-component plasma (OCP), the random-phase approximation (RPA) gives an equal-time potential correlator that is finite at short distance but retains an unscreened $1/r$ tail, as expected from classical plasma fluctuation theory~\cite{klimontovich1986statistical}. In our preceding quasi-static analysis~\cite{budkov2026static}, this tail generated an infrared Coulomb logarithm in the eikonal attenuation length of the averaged retarded Green function. The weak and formal strong saddles were shown to be distinguished by the dimensionless parameter $g=G/k^3$, whereas the straight-line approximation requires an independent separation between the particle wavelength and the spatial correlation scale. The same work treated mutual coherence in a slab of finite thickness $d$. Its central conclusion was that the phase-difference structure function is finite: an algebraic coherence envelope exists only as an intermediate asymptotic, and the structure function saturates for transverse separations larger than the slab thickness. Consequently, a transverse $e^{-1}$ coherence radius is conditional rather than universal.

The quasi-static approximation is appropriate when the ionic configuration changes little during the passage of the test particle. A real OCP, however, supports ion plasma oscillations and ballistic thermal motion. When the test-particle velocity is comparable to or smaller than the ion thermal speed, the fluctuating Coulomb field cannot be treated as frozen. The temporal decorrelation of long-wavelength modes then modifies the line-integrated disorder sampled along the trajectory and can remove the infrared logarithm of the static problem.

A complementary density-matrix analysis of Coulomb environments~\cite{budkov2026annals} distinguishes the random-unitary evolution generated by a prescribed classical field from the Born--Redfield dynamics of an equilibrium quantum bath. In the latter case, ordered bath correlators give rise to detailed balance and, in the kinetic limit, to momentum diffusion and friction. That analysis also shows that momentum-transition rates do not uniquely determine the evolution of quantum coherences. The present work examines phase coherence within a more specific setting: a classical Gaussian Coulomb potential with a temporal spectrum determined by equilibrium ionic fluctuations. We calculate the averaged propagator and mutual coherence directly from its space--time covariance, without reconstructing them from a population kinetic equation or assuming a complete quantum-bath description.

Recent electromagnetic Green-tensor approaches have examined free-electron decoherence in bulk materials and near surfaces~\cite{velasco2026free}, and in cylindrical holes and circular apertures~\cite{velasco2026apertures}. These works emphasize the material response and sample geometry. Here we isolate classical ionic charge fluctuations and determine how collisionless and diffusive relaxation affect the attenuation of an averaged electronic amplitude and the spatial coherence of two trajectories.

We now extend the quasi-static calculation of Ref.~\cite{budkov2026static} to temporally correlated ionic fluctuations, retaining its finite-size assumptions and physical interpretation of the attenuation length. We first derive the averaged propagator for a stationary Gaussian time-dependent potential and identify the dynamic line-integrated disorder strength. We then obtain the OCP potential spectrum from the fluctuation--dissipation theorem and the RPA longitudinal dielectric function. The fast- and slow-particle limits are evaluated analytically. Particular attention is paid to the saddle variable: away from the weak branch, the correlation function is sampled at the effective velocity $v_s=v/s$, so the on-shell result cannot simply be inserted unchanged into the static saddle equation.

Mutual coherence is treated from the exact finite-thickness phase structure function rather than from an infinite-path approximation. In the slow weak-saddle regime the thick-slab structure function can be evaluated for arbitrary transverse separation, yielding a closed scaling function and an explicit saturation plateau. To separate the effect of static Coulomb correlations from that of charge relaxation, we also consider a symmetric binary electrolyte with overdamped diffusive ions. Linearized stochastic density functional theory (SDFT) gives a Debye--Falkenhagen charge mode and permits an exact evaluation of the dynamic line strength. The experimentally relevant limits differ between the two media. In suitable collisionless plasmas both the fast and slow probe limits may be accessible, whereas experimentally realistic coherent, spatially eikonal probes in ordinary liquid electrolytes lie in the fast-diffusion limit. In the latter case the central new prediction is an intrinsic dynamical infrared cutoff, replacing the externally imposed cutoff of the quasi-static calculation when the sample geometry allows it to become effective. The slow-electrolyte scaling functions are retained as analytically well-defined model limits, not as direct predictions for conventional liquid-cell experiments.

\section{Averaged propagator in a dynamic Gaussian potential}
\label{sec:prelim}

We first derive the propagation problem for a general stationary Gaussian potential. This identifies the line-integrated correlations sampled along an approximately rectilinear trajectory before we specify the ionic dynamics.

We consider a nonrelativistic quantum particle of mass $m$ moving in a stationary time-dependent random potential energy $W(\mathbf{x},t)$, with $H(t)=\mathbf{p}^2/(2m)+W(\mathbf{x},t)$. The retarded propagator is
\begin{equation}\label{eq:propagator_def}
\begin{aligned}
G(\mathbf{x},t;\mathbf{y},0)
={}&\theta(t)\left\langle\mathbf{x}\left|
\mathcal T\exp\left[-\frac{i}{\hbar}\int_0^t H(\tau)d\tau\right]
\right|\mathbf y\right\rangle .
\end{aligned}
\end{equation}
Here $\mathcal T$ denotes chronological time ordering: in the Dyson expansion,
operators with later time arguments are placed to the left. This ordering is
required because $H(t)$ at different times need not commute. In the
coordinate-space path integral below it is already encoded by the usual
time-slicing prescription and therefore does not appear as an additional
operator.
Its path-integral representation reads~\cite{zinn2010path}
\begin{equation*}
\begin{aligned}
G(\mathbf{x},t;\mathbf{y},0)
={}&\theta(t)\int_{\mathbf y}^{\mathbf x}\mathcal D\mathbf x(\tau)\exp\left[\frac{im}{2\hbar}\int_0^t d\tau\,
\dot{\mathbf x}^{2}(\tau)\right]\\
&\times\exp\left[-\frac{i}{\hbar}\int_0^t d\tau\,
W(\mathbf x(\tau),\tau)\right].
\end{aligned}
\end{equation*}
After ensemble averaging, stationarity restores dependence only on the time difference. The energy-resolved averaged propagator is defined by
\begin{equation*}
\begin{aligned}
\left\langle G_k(\mathbf{x},\mathbf{y})\right\rangle
={}&-\frac{i}{\hbar}\int_0^\infty dt\,
\exp\left[\frac{i}{\hbar}(E+i0)t\right]\\
&\times\left\langle G(\mathbf{x},t;\mathbf{y},0)\right\rangle,
\qquad E=\frac{\hbar^2k^2}{2m}.
\end{aligned}
\end{equation*}

Following Efimov~\cite{efimov2015quantum}, we introduce
\begin{equation*}
\begin{aligned}
t&=\frac{mr}{\hbar k}s,\qquad
\tau=\frac{ms}{\hbar k}u,\\
\boldsymbol\xi(\tau)&=\sqrt{\frac{s}{k}}\,\boldsymbol\eta(u).
\end{aligned}
\end{equation*}
where $r=|\mathbf{x}-\mathbf{y}|$, $\mathbf{n}=(\mathbf{x}-\mathbf{y})/r$, $u\in[0,r]$, and $\boldsymbol\eta(0)=\boldsymbol\eta(r)=0$. It is convenient to denote the rescaled path and its physical time by $\mathbf x_s(u)=\mathbf y+\mathbf n u-\sqrt{s/k}\,\boldsymbol\eta(u)$ and $t_s(u)=msu/(\hbar k)$. Up to normalization constants that do not affect the decay exponent,
\begin{equation}\label{eq:Efimov_dynamic}
\begin{aligned}
G_k(\mathbf{x},\mathbf{y}|W)
={}&B\int_0^\infty\frac{ds}{s^{3/2}}\exp\left[\frac{i}{2}kr\left(s+\frac{1}{s}\right)\right]\\
&\times\int\frac{\mathcal D\boldsymbol\eta}{\mathcal C}\exp\left[\frac{i}{2}\int_0^rdu\,
\dot{\boldsymbol\eta}^{2}(u)\right]\\
&\times\exp\left[-i\frac{ms}{\hbar^2k}\int_0^rdu\,
W\!\left(\mathbf x_s(u),t_s(u)\right)\right].
\end{aligned}
\end{equation}

The random potential energy is taken to be a zero-mean Gaussian field with $\langle W(\mathbf{x},t)W(\mathbf{x}',t')\rangle=K(\mathbf{x}-\mathbf{x}',t-t')$. In the double integral below we use $\Delta\mathbf x_s=\mathbf x_s(u)-\mathbf x_s(u')$ and $\Delta t_s=t_s(u)-t_s(u')$. Gaussian averaging gives
\begin{equation}\label{eq:averaged_dynamic_G}
\begin{aligned}
\left\langle G_k(r)\right\rangle
={}&B\int_0^\infty\frac{ds}{s^{3/2}}
\int\frac{\mathcal D\boldsymbol\eta}{\mathcal C}\\
&\times\exp\left[\frac{i}{2}kr\left(s+\frac{1}{s}\right)\right]\\
&\times\exp\left[\frac{i}{2}\int_0^rdu\,
\dot{\boldsymbol\eta}^{2}(u)\right]\\
&\times\exp\left[-\frac{m^2s^2}{2\hbar^4k^2}
\int_0^rdu\int_0^rdu'\,
K\!\left(\Delta\mathbf x_s,\Delta t_s\right)\right].
\end{aligned}
\end{equation}

In the straight-line approximation, $\boldsymbol\eta=0$. For a real positive $s$ and a propagation distance large compared with the line-correlation range, the exponent per unit length is
\begin{equation*}
\begin{aligned}
A(s;k)
&=-\frac{m^2s^2}{2\hbar^4k^2}
\int_{-\infty}^{\infty}du\,
K\left(\mathbf n u,\frac{u}{v_s}\right),\\
v_s&=\frac{\hbar k}{ms}=\frac{v}{s}.
\end{aligned}
\end{equation*}
Here $v=\hbar k/m$. We define the on-shell dynamic line strength for a real positive velocity by
\begin{equation}\label{eq:Gdyn_definition}
\begin{aligned}
G(v)
&=\frac{m^2}{2\hbar^4}
\int_{-\infty}^{\infty}du\,
K\left(\mathbf n u,\frac{u}{v}\right)\\
&=\frac{m^2}{\hbar^4}
\int_0^\infty du\,
K\left(\mathbf n u,\frac{u}{v}\right).
\end{aligned}
\end{equation}
Thus, on the positive real $s$ axis,
\begin{equation}\label{eq:A_s_dynamic}
A(s;k)=-\frac{s^2}{k^2}G\left(\frac{v}{s}\right).
\end{equation}
The saddle exponent is
\begin{equation}\label{eq:Phi_general}
\Phi(s;k)=\frac{ik}{2}\left(s+\frac{1}{s}\right)+A(s;k),
\qquad
\partial_s\Phi(s_c;k)=0,
\end{equation}
and the eikonal attenuation rate is
\begin{equation}\label{eq:Gamma_definition}
\Gamma(k)=\frac{1}{\ell(k)}=-\operatorname{Re}\Phi(s_c;k).
\end{equation}
The physical meaning of $\ell$ is most transparent in the controlled weak on-shell branch. After ensemble averaging restores translational invariance, one may write $\overline{G^R}(\mathbf p,E)=[E-\varepsilon_p-\Sigma^R(\mathbf p,E)]^{-1}$. If $\Sigma^R=\Delta-i\Gamma_\Sigma$ varies slowly across the spectral line, displacement of the on-shell pole gives
\begin{equation}\label{eq:ell_spectral_meaning}
\ell=\frac{\hbar v_g}{\Gamma_\Sigma}
=\frac{\hbar v_g}{-\operatorname{Im}\Sigma^R_{\rm on\,shell}},
\end{equation}
where $v_g$ is the renormalized group velocity~\cite{abrikosov1988fundamentals}. Thus $\ell$ is the spatial $e$-folding length of the coherent ensemble-averaged one-particle amplitude, or equivalently the spatial counterpart of its spectral linewidth. It is not a transport mean free path, an intensity attenuation length, or an Anderson localization length. In the Gaussian eikonal language the same statement follows from $\overline{G^R}/G_0^R=\exp[-\langle\phi^2\rangle/2]$: at $r=\ell$ the accumulated phase variance is of order unity, $\langle\phi^2\rangle=2$.

For a complex saddle, Eq.~\eqref{eq:A_s_dynamic} requires analytic continuation of the line-correlation integral. The on-shell real-velocity formula derived below is therefore automatically controlled only in the weak branch $s_c\simeq1$. Any strong-saddle continuation must be assessed separately.

\section{Dynamic potential correlations in a classical OCP}
\label{sec:OCP_dynamic}

We consider point ions of charge $q$, mass $m_i$, and mean number density $n_0$ in a uniform neutralizing background. SI electrostatic units are used throughout. The absolute background permittivity is denoted by $\varepsilon_b=\varepsilon_0\varepsilon_{b,r}$. It corresponds to $\varepsilon$ in the quasi-static theory of Ref.~\cite{budkov2026static}. The subscript distinguishes it from the longitudinal dielectric function $\varepsilon(Q,\omega)$. The microscopic charge-density fluctuation is $\delta\rho_c(\mathbf r,t)=q\sum_j\delta(\mathbf r-\mathbf r_j(t))-qn_0$, with Fourier convention $\delta\rho_c(\mathbf Q,t)=\int d^3r\,e^{-i\mathbf Q\cdot\mathbf r}\delta\rho_c(\mathbf r,t)$.
The fluctuating electrostatic potential energy of a test particle of charge $q_0$ is
\begin{equation}\label{eq:W_from_charge}
W(\mathbf{Q},t)=\frac{q_0}{\varepsilon_b Q^2}\,\delta\rho_c(\mathbf{Q},t).
\end{equation}

Here the OCP is treated specifically as a collisionless classical ionic
plasma with Maxwellian ion velocities. This provides an analytically
tractable reference model for dynamic Coulomb fluctuations. Other ionic
media need not share this microscopic dynamics. In particular, the
diffusive electrolyte considered below has the same Gaussian equal-time
Coulomb correlations but a different temporal relaxation law.

\subsection{RPA response and fluctuation--dissipation relation}

We first define the charge-density spectrum in a periodic volume $V$ and take the thermodynamic limit afterward. With the Fourier convention stated above, the intensive spectrum is
\begin{equation*}
\widetilde S_{\rho\rho}(Q,\omega)
=\lim_{V\to\infty}\frac{1}{V}
\int_{-\infty}^{\infty}dt\,e^{i\omega t}
\left\langle\delta\rho_c(\mathbf{Q},t)\delta\rho_c(-\mathbf{Q},0)\right\rangle.
\end{equation*}
In continuum notation, the corresponding two-point function carries the factor $(2\pi)^3\delta(\mathbf Q+\mathbf Q')$. Throughout this paper, spatial spectra denote the coefficients of this delta function, not the full Fourier-mode expectation values.
The potential-energy spectrum is therefore $\widetilde K(Q,\omega)=[q_0/(\varepsilon_bQ^2)]^2\widetilde S_{\rho\rho}(Q,\omega)$. We define the retarded charge response by $\langle\delta\rho_c(\mathbf Q,\omega)\rangle=\chi_{\rho\rho}(Q,\omega)\phi_{\rm ext}(\mathbf Q,\omega)$ for a perturbation $H'=\int d^3r\,\delta\rho_c\phi_{\rm ext}$. With this convention the static susceptibility is negative. In the RPA,
\begin{equation*}
\chi_{\rho\rho}(Q,\omega)
=\frac{\chi_0(Q,\omega)}
{1-\frac{1}{\varepsilon_b Q^2}\chi_0(Q,\omega)},
\end{equation*}
where, for a Maxwellian ion plasma~\cite{pitaevskii2012physical,fried1961plasma},
\begin{equation*}
\begin{aligned}
\chi_0(Q,\omega)
&=-\frac{n_0q^2}{k_{\mathrm B}T}\left[1+\xi Z(\xi)\right],\\
\xi&=\frac{\omega}{\sqrt{2}Qv_{\rm th}},\qquad
v_{\rm th}=\sqrt{\frac{k_{\mathrm B}T}{m_i}}.
\end{aligned}
\end{equation*}
Here $Z(\xi)$ is the plasma dispersion function, defined by $Z(\xi)=\pi^{-1/2}\int_{-\infty}^{\infty}dt\,e^{-t^2}/(t-\xi)$ for $\operatorname{Im}\xi>0$ and continued analytically with the Landau prescription. The Boltzmann constant is denoted by $k_{\mathrm B}$, and $T$ is the temperature.
The longitudinal dielectric function is
\begin{equation}\label{eq:longitudinal_dielectric}
\begin{aligned}
\varepsilon(Q,\omega)
&=1-\frac{1}{\varepsilon_b Q^2}\chi_0(Q,\omega)\\
&=1+\frac{\kappa^2}{Q^2}\left[1+\xi Z(\xi)\right],\qquad
\kappa^2=\frac{n_0q^2}{\varepsilon_b k_{\mathrm B}T}.
\end{aligned}
\end{equation}
The corresponding Debye length is $\lambda_D=\kappa^{-1}$. It follows that $\chi_{\rho\rho}(Q,\omega)=\varepsilon_b Q^2[\varepsilon^{-1}(Q,\omega)-1]$.

In the classical limit the fluctuation--dissipation theorem gives $\widetilde S_{\rho\rho}(Q,\omega)=-(2k_{\mathrm B}T/\omega)\operatorname{Im}\chi_{\rho\rho}(Q,\omega)$. Consequently,
\begin{equation}\label{eq:dynamic_potential_spectrum}
\widetilde K(Q,\omega)
=-\frac{2 q_0^2k_{\mathrm B}T}
{\varepsilon_b Q^2\omega}
\operatorname{Im}\frac{1}{\varepsilon(Q,\omega)}.
\end{equation}
For an equilibrium classical density field, $\widetilde K(Q,\omega)$ is nonnegative and even in $\omega$.

\subsection{Equal-time check}

The equal-time spectrum is $K(Q)=\int_{-\infty}^{\infty}(d\omega/2\pi)\widetilde K(Q,\omega)$. For $f(Q,\omega)=1/\varepsilon(Q,\omega)-1$, the Kramers--Kronig relation gives
\begin{equation*}
\begin{aligned}
\frac{1}{\pi}\int_{-\infty}^{\infty}
\frac{d\omega}{\omega}
\operatorname{Im}\frac{1}{\varepsilon(Q,\omega)}
&=\frac{1}{\varepsilon(Q,0)}-1\\
&=-\frac{\kappa^2}{Q^2+\kappa^2}.
\end{aligned}
\end{equation*}
Substitution into Eq.~\eqref{eq:dynamic_potential_spectrum} yields
\begin{equation}\label{eq:static_spectrum_recovered}
K(Q)=\frac{q_0^2k_{\mathrm B}T}{\varepsilon_b}
\frac{\kappa^2}{Q^2(Q^2+\kappa^2)}.
\end{equation}
Fourier inversion gives
\begin{equation}\label{eq:static_correlator_recovered}
K(r)=C\frac{1-e^{-\kappa r}}{r},
\qquad
C=\frac{k_{\mathrm B}Tq_0^2}{4\pi\varepsilon_b},
\end{equation}
which is exactly the quasi-static RPA correlator of Ref.~\cite{budkov2026static}.

\subsection{Dynamic line strength}

For a real on-shell velocity $v$, the line correlator is
\begin{equation*}
\begin{aligned}
K_{\rm dyn}(u)
={}&K\left(\mathbf n u,\frac{u}{v}\right)\\
={}&\int\frac{d^3Q}{(2\pi)^3}
\int_{-\infty}^{\infty}\frac{d\omega}{2\pi}\,
\widetilde K(Q,\omega)\\
&\times e^{i(Q_\parallel-\omega/v)u}.
\end{aligned}
\end{equation*}
Using $\int_0^\infty du\,e^{i\alpha u}=\pi\delta(\alpha)+i\mathcal P(1/\alpha)$, the real attenuation exponent is determined by the resonant delta-function part. The result is
\begin{equation}\label{eq:Gdyn_spectral}
G(v)
=\frac{m^2}{8\pi^2\hbar^4}
\int_0^\infty dQ\,Q
\int_{-Qv}^{Qv}d\omega\,\widetilde K(Q,\omega).
\end{equation}
Equivalently,
\begin{equation}\label{eq:Gdyn_master}
\begin{aligned}
G(v)
={}&-\frac{m^2q_0^2k_{\mathrm B}T}
{2\pi^2\varepsilon_b\hbar^4}
\int_0^\infty\frac{dQ}{Q}\int_0^{Qv}\frac{d\omega}{\omega}
\operatorname{Im}\frac{1}{\varepsilon(Q,\omega)}.
\end{aligned}
\end{equation}
Equation~\eqref{eq:Gdyn_master} is the central on-shell formula for the dynamic attenuation problem.

\subsection{Fast-particle limit}

For $v\gg v_{\rm th}$, the Doppler window $|\omega|<Qv$ contains essentially the full ionic spectrum except at sufficiently small $Q$. The long-wavelength response is controlled by $\omega_{\rm p}^2=n_0q^2/(\varepsilon_b m_i)=\kappa^2v_{\rm th}^2$. Thus the dynamic decorrelation imposes a lower wave number of order $\omega_{\rm p}/v$. A separate physical infrared cutoff $L_{\rm IR}$, already required in the quasi-static averaged propagator, must also be retained. To logarithmic accuracy,
\begin{equation}\label{eq:effective_IR_length}
\begin{aligned}
Q_{\min}&\sim\max\left(L_{\rm IR}^{-1},\frac{\omega_{\rm p}}{v}\right),\\
L_{\rm eff}&=Q_{\min}^{-1}
=\min\left(L_{\rm IR},\frac{v}{\omega_{\rm p}}\right).
\end{aligned}
\end{equation}
Using the static Kramers--Kronig integral for modes with $Q\gtrsim Q_{\min}$ gives
\begin{equation}\label{eq:G_fast}
\begin{aligned}
G_{\rm fast}(v)
&\simeq\frac{m^2C}{\hbar^4}
\int_{Q_{\min}}^\infty
\frac{\kappa^2\,dQ}{Q(Q^2+\kappa^2)}
\\
&=\frac{m^2C}{2\hbar^4}
\ln\left[1+(\kappa L_{\rm eff})^2\right]
\simeq\frac{m^2C}{\hbar^4}\ln(\kappa L_{\rm eff}),
\end{aligned}
\end{equation}
where $C$ is defined in Eq.~\eqref{eq:static_correlator_recovered}, and the last form assumes $\kappa L_{\rm eff}\gg1$. If $v/\omega_{\rm p}<L_{\rm IR}$, the logarithm is $\ln(v/v_{\rm th})$ within logarithmic accuracy. If $v/\omega_{\rm p}>L_{\rm IR}$, the finite physical cutoff of the static problem is recovered.

\subsection{Slow-particle limit}

For $v\ll v_{\rm th}$, the sampled frequencies satisfy $|\omega|/(Qv_{\rm th})\ll1$. The plasma dispersion function has the expansion $Z(\xi)=i\sqrt{\pi}-2\xi-i\sqrt{\pi}\xi^2+\frac{4}{3}\xi^3+O(\xi^4)$. The quadratic term in $Z(\xi)$ contributes only at cubic order to the imaginary part of the dielectric function. The leading frequency-dependent loss is therefore
\begin{equation}\label{eq:low_frequency_loss}
\begin{aligned}
\operatorname{Im}\frac{1}{\varepsilon(Q,\omega)}
={}&-\sqrt{\frac{\pi}{2}}
\frac{\kappa^2Q}{(Q^2+\kappa^2)^2}
\frac{\omega}{v_{\rm th}}\\
&+O\left(\frac{\omega^3}{Q^3v_{\rm th}^3}\right).
\end{aligned}
\end{equation}
Substitution into Eq.~\eqref{eq:Gdyn_master} yields
\begin{equation}\label{eq:G_slow}
G_{\rm slow}(v)
=\frac{1}{\sqrt{2\pi}}
\frac{m^2C}{\hbar^4}\frac{v}{v_{\rm th}}
+O\left(\frac{v^3}{v_{\rm th}^3}\right).
\end{equation}
The infrared logarithm has disappeared. Temporal decorrelation renders the line integral finite and makes the effective line strength vanish linearly as $v\to0$.

\section{Weak and formal strong dynamic saddles}
\label{sec:dynamic_saddles}

As in the quasi-static theory, we define the dimensionless saddle parameter
$g(v)=G(v)/k^3$. The weak branch corresponds to $g(v)\ll1$. When this
condition is satisfied, $s_c\simeq1$, and the on-shell line strength gives
\begin{equation}\label{eq:weak_dynamic_attenuation}
\Gamma(k)=\frac{G(v)}{k^2},
\qquad
\ell(k)=\frac{k^2}{G(v)}.
\end{equation}
This saddle condition is distinct from the eikonal requirement. A sufficient spatial condition for the OCP model is $k\gg\kappa$, together with a propagation distance large compared with the microscopic wavelength.

In the fast regime, $\ell_{\rm fast}(k)\simeq\hbar^4k^2/[m^2C\ln(\kappa L_{\rm eff})]$ for $v\gg v_{\rm th}$ and $g(v)\ll1$. In the slow regime, $v=\hbar k/m$ and Eq.~\eqref{eq:G_slow} give $G(v)=g_0k$, with $g_0=mC/(\sqrt{2\pi}\hbar^3v_{\rm th})$, so that
\begin{equation}\label{eq:ell_slow_weak}
\ell_{\rm slow}(k)
=\frac{k}{g_0}
=\sqrt{2\pi}\,
\frac{\hbar^3v_{\rm th}}{mC}\,k,
\qquad
\frac{g_0}{k^2}\ll1.
\end{equation}
Thus the dynamic weak-branch attenuation length grows linearly with $k$, rather than quadratically. Although $G\to0$ as $v\to0$, the weak-branch rate $G/k^2=g_0/k$ increases as $k$ decreases. Temporal decorrelation weakens this growth relative to the quasi-static $1/k^2$ law but does not remove it.

The strong-saddle continuation requires the $s$ dependence of the sampled velocity. If the slow-response form in Eq.~\eqref{eq:G_slow} is analytically continued along the relevant saddle contour, then $G(v/s)=g_0k/s$, and
\begin{equation*}
\Phi_{\rm slow}(s;k)
=\frac{ik}{2}\left(s+\frac{1}{s}\right)-\frac{g_0 s}{k}.
\end{equation*}
The saddle connected continuously to $s=1$ is $s_c=(1+2ig_0/k^2)^{-1/2}$, at which $\Phi_{\rm slow}(s_c;k)=ik\sqrt{1+2ig_0/k^2}$. Accordingly, the formal slow-response interpolation for the attenuation rate is
\begin{equation}\label{eq:slow_saddle_rate}
\Gamma_{\rm slow}(k)
=k\left[
\frac{\sqrt{1+4g_0^2/k^4}-1}{2}
\right]^{1/2}.
\end{equation}
Its limits are
\begin{equation}\label{eq:slow_saddle_limits}
\begin{aligned}
\ell_{\rm slow}(k)=\Gamma_{\rm slow}^{-1}(k)&\simeq
\begin{cases}
k/g_0, & g_0/k^2\ll1,\\[2pt]
g_0^{-1/2}, & g_0/k^2\gg1.
\end{cases}
\end{aligned}
\end{equation}
The momentum-independent second line is a formal strong-saddle saturation, not a proof of Anderson localization or pinning. Its use requires all of the following: analytic continuation of the slow-response line integral, a saddle-sampled velocity satisfying $|v/s_c|\ll v_{\rm th}$, and an independent straight-line condition such as $k\gg\kappa$. In the strong limit $|v/s_c|\to\hbar\sqrt{2g_0}/m$, so the slow-response condition becomes a constraint on the plasma and test-particle parameters rather than on $k$ alone. There need not be a parametrically broad domain in which every condition is simultaneously satisfied.

\section{Finite-thickness mutual coherence}
\label{sec:dynamic_coherence}

This section turns from attenuation of a single ensemble-averaged amplitude to the phase difference between two transmitted amplitudes. As in the quasi-static theory, sufficiently long-wavelength fluctuations that are common to both paths cancel from this two-ray observable, so its infrared structure need not follow that of the one-particle propagator.

We consider two parallel trajectories of length $d$ separated by a fixed transverse vector $\boldsymbol\rho$. As in the quasi-static theory, the eikonal phase coefficient is $a=m/(k\hbar^2)=1/(\hbar v)$. The normalized mutual coherence is
\begin{equation}\label{eq:coherence_definition}
\gamma_d(\rho)=\exp\left[-\frac{1}{2}\mathcal D_d(\rho)\right].
\end{equation}
This factor is the straight-ray retarded--advanced measure of mutual coherence.

\subsection{Exact finite-slab structure function}

For the stationary classical random field, with $\Delta z=z-z'$,
\begin{equation*}
\begin{aligned}
\mathcal D_d(\rho)
=2a^2\int_0^d dz\int_0^d dz'\,
\bigg[&K\left(\mathbf n\Delta z,\frac{\Delta z}{v}\right)\\
&-K\left(\boldsymbol\rho+\mathbf n\Delta z,
\frac{\Delta z}{v}\right)\bigg].
\end{aligned}
\end{equation*}
Equivalently, introducing $F_d(\Omega)=\int_0^d dz\,e^{i\Omega z}=e^{i\Omega d/2}2\sin(\Omega d/2)/\Omega$, we obtain the manifestly nonnegative spectral representation
\begin{equation}\label{eq:Dphi_finite_spectral}
\begin{aligned}
\mathcal D_d(\rho)
=2a^2\int\frac{d^3Q}{(2\pi)^3}
\int_{-\infty}^{\infty}\frac{d\omega}{2\pi}\,
&\left[1-\cos(\mathbf{Q}_\perp\cdot\boldsymbol\rho)\right]
\widetilde K(Q,\omega)
\\
&\times\left|F_d\left(Q_\parallel-\frac{\omega}{v}\right)\right|^2.
\end{aligned}
\end{equation}
Equation~\eqref{eq:Dphi_finite_spectral}, rather than an infinite-path expression, is the correct starting point for a slab of finite thickness. In particular,
\begin{equation*}
\begin{aligned}
\mathcal D_d(\infty)
={}&2a^2\int\frac{d^3Q}{(2\pi)^3}
\int_{-\infty}^{\infty}\frac{d\omega}{2\pi}\,
\widetilde K(Q,\omega)\\
&\times\left|F_d\left(Q_\parallel-\frac{\omega}{v}\right)\right|^2.
\end{aligned}
\end{equation*}
provided the transverse Fourier transform is integrable. Since $\mathcal D_d(0)=0$ and the structure function is continuous, the plateau condition
\begin{equation}\label{eq:coherence_existence_general}
\mathcal D_d(\infty)>2
\end{equation}
is sufficient for at least one finite coherence radius. For a general dynamic spectrum it is not necessary if $\mathcal D_d(\rho)$ is nonmonotonic. For the monotone slow-OCP and diffusive-electrolyte cases derived below, Eq.~\eqref{eq:coherence_existence_general} is both necessary and sufficient. Equality corresponds to the marginal case $\rho_c\to\infty$.

If $d$ is much larger than the effective line-correlation range, then $|F_d(\Omega)|^2\to2\pi d\,\delta(\Omega)$, and
\begin{equation}\label{eq:Dphi_thick_general}
\begin{aligned}
\mathcal D_d(\rho)
\simeq{}&2a^2dv\int\frac{d^3Q}{(2\pi)^3}
\left[1-\cos(\mathbf Q_\perp\cdot\boldsymbol\rho)\right]\\
&\times\widetilde K(Q,vQ_\parallel).
\end{aligned}
\end{equation}
For an integrable dynamic line correlator, the thick-slab plateau is directly related to the weak on-shell attenuation length:
\begin{equation}\label{eq:plateau_ell_relation}
\mathcal D_d(\infty)
\simeq4d\frac{G(v)}{k^2}
=\frac{4d}{\ell(k)}.
\end{equation}
This relation involves $\ell=k^2/G(v)$. It does not identify the mutual-coherence plateau with the formal strong-saddle length of Eq.~\eqref{eq:slow_saddle_limits}.

\subsection{Closed slow-limit scaling function}

In the slow regime, Eq.~\eqref{eq:low_frequency_loss} gives the sampled potential spectrum
\begin{equation*}
\begin{aligned}
\widetilde K(Q,vQ_\parallel)
={}&8\pi\sqrt{\frac{\pi}{2}}
\frac{C\kappa^2}{v_{\rm th}}
\frac{1}{Q(Q^2+\kappa^2)^2}\\
&+O\left(\frac{v^2}{v_{\rm th}^2}\right).
\end{aligned}
\end{equation*}
Substitution into Eq.~\eqref{eq:Dphi_thick_general} and angular integration yield
\begin{equation}\label{eq:Dphi_slow_scaling}
\mathcal D_d(\rho)
\simeq8\frac{d}{\ell_{\rm slow}}\,
\mathcal F(\kappa\rho),
\end{equation}
where
\begin{equation}\label{eq:J_integral}
\mathcal F(x)=\int_0^\infty dy\,
\frac{y}{(1+y^2)^2}
\left[1-\frac{\sin(xy)}{xy}\right].
\end{equation}
For $x>0$, this function can be written as
\begin{equation*}
\mathcal F(x)
=\frac{1}{2}
-\frac{1}{4x}\left[
(x-1)e^x\operatorname{Ei}(-x)
+(x+1)e^{-x}\operatorname{Ei}(x)
\right].
\end{equation*}
The small- and large-$x$ limits are
\begin{equation}\label{eq:J_small}
\begin{aligned}
\mathcal F(x)
={}&\frac{x^2}{6}
\left[\ln\frac{1}{x}-\gamma_{\rm E}+\frac{4}{3}\right]
+O\left(x^4\ln x\right),\\
&x\ll1,
\end{aligned}
\end{equation}
\begin{equation*}
\mathcal F(x)
=\frac{1}{2}-\frac{1}{x^2}+O(x^{-4}),
\qquad x\gg1.
\end{equation*}
Consequently,
\begin{equation}\label{eq:Dphi_slow_small}
\begin{aligned}
\mathcal D_d(\rho)
\simeq{}&\frac{4}{3}\frac{d\kappa^2\rho^2}{\ell_{\rm slow}}
\left[\ln\frac{1}{\kappa\rho}-\gamma_{\rm E}+\frac{4}{3}\right],\\
&\kappa\rho\ll1,
\end{aligned}
\end{equation}
whereas
\begin{equation*}
\begin{aligned}
\mathcal D_d(\rho)
\simeq{}&\frac{4d}{\ell_{\rm slow}}
\left[1-\frac{2}{(\kappa\rho)^2}+O((\kappa\rho)^{-4})\right],\\
&\kappa\rho\gg1.
\end{aligned}
\end{equation*}
Temporal decorrelation therefore removes the static logarithmic growth at large transverse separation. In the slow thick-slab regime the mutual coherence approaches the nonzero plateau $\gamma_d(\infty)=\exp(-2d/\ell_{\rm slow})$.

\subsection{Conditional coherence radius}

The closed scaling function $\mathcal F(x)$ increases from $0$ to $1/2$. Within the thick-slab slow approximation, a finite radius therefore exists if and only if $d>\ell_{\rm slow}/2$. Equality is marginal. If this condition is not met, the dynamic ionic fluctuations reduce the mutual coherence by a finite amount but do not define a transverse coherence radius.

If a radius $\rho_c$ exists and also satisfies $\kappa\rho_c\ll1$, Eq.~\eqref{eq:Dphi_slow_small} gives
\begin{equation}\label{eq:rho_c_conditional}
\begin{aligned}
(\kappa\rho_c)^2
\left[\ln\frac{1}{\kappa\rho_c}-\gamma_{\rm E}+\frac{4}{3}\right]
=\frac{3}{2}\frac{\ell_{\rm slow}}{d}.
\end{aligned}
\end{equation}
Only under these conditions can one summarize the result parametrically as
\begin{equation}\label{eq:rho_c_parametric}
\rho_c\sim\lambda_D
\sqrt{\frac{\ell_{\rm slow}}{d}},
\end{equation}
with a logarithmic correction determined self-consistently by Eq.~\eqref{eq:rho_c_conditional}. Equation~\eqref{eq:rho_c_parametric} is not a universal relation and must not be extrapolated to the formal strong saddle or to a slab whose plateau remains above $e^{-1}$.

\section{Diffusive electrolyte: exact SDFT result}
\label{sec:sdftelectrolyte}

The collisionless OCP provides one realization of a time-dependent Coulomb environment. In a dilute electrolyte, by contrast, the long-wavelength ionic dynamics is overdamped and diffusive. It is therefore useful to ask which results above follow from the common static Debye--H\"uckel correlations and which depend on the dynamical model. We consider a symmetric $1{:}1$ electrolyte with ionic charges $q_\pm=\pm q$, equal mean concentrations $n_s$, and a common diffusion coefficient $D$. The linearized SDFT used below is the Gaussian limit of the Dean--Kawasaki description and is standard in field-theoretic treatments of electrolyte dynamics~\cite{demery2016conductivity,avni2022conductivity,illien2024sdftelectrolyte}.

At the mean-field electrostatic level we use the point-ion free-energy functional, with $\rho_c=q(n_+-n_-)$,
\begin{equation*}
\begin{aligned}
\beta F[n_+,n_-]
={}&\sum_{\alpha=\pm}\int d^3r\,
n_\alpha(\mathbf r)
\left[\ln\frac{n_\alpha(\mathbf r)}{n_s}-1\right]\\
&+\frac{\beta}{8\pi\varepsilon_b}
\int d^3r\,d^3r'\,
\frac{\rho_c(\mathbf r)\rho_c(\mathbf r')}
{|\mathbf r-\mathbf r'|},
\end{aligned}
\end{equation*}
where $\beta=(k_{\text{B}} T)^{-1}$.
The corresponding SDFT equations are
\begin{equation}\label{eq:SDFT_general}
\begin{aligned}
\partial_t n_\alpha
={}&D\nabla\cdot\left[
n_\alpha\nabla\frac{\delta \beta F}{\delta n_\alpha}
\right]\\
&+\nabla\cdot\left[\sqrt{2D n_\alpha}\,
\boldsymbol\Lambda_\alpha\right],\qquad \alpha=\pm,
\end{aligned}
\end{equation}
with
\begin{equation*}
\begin{aligned}
\left\langle
\Lambda_{\alpha i}(\mathbf r,t)
\Lambda_{\beta j}(\mathbf r',t')
\right\rangle
={}&\delta_{\alpha\beta}\delta_{ij}\\
&\times\delta(\mathbf r-\mathbf r')\delta(t-t').
\end{aligned}
\end{equation*}
Linearization about $n_\pm=n_s$ separates the total-density and charge modes. For $\delta\rho_c=q(\delta n_+-\delta n_-)$ the charge mode obeys
\begin{equation}\label{eq:SDFT_charge_mode}
\begin{aligned}
\partial_t\delta\rho_c(\mathbf r,t)
={}&D\left(\nabla^2-\kappa^2\right)\delta\rho_c(\mathbf r,t)
+\nabla\cdot\boldsymbol\eta_c(\mathbf r,t),\\
&\kappa^2=\frac{2n_s q^2}{\varepsilon_b k_{\mathrm B}T}.
\end{aligned}
\end{equation}
The noise amplitude is fixed by equilibrium fluctuation--dissipation balance. The stationary charge correlation function is
\begin{equation}\label{eq:SDFT_charge_corr}
\begin{aligned}
\frac{1}{V}\left\langle
\delta\rho_c(\mathbf Q,t)\delta\rho_c(-\mathbf Q,0)
\right\rangle
={}&\varepsilon_b k_{\mathrm B}T
\frac{\kappa^2Q^2}{Q^2+\kappa^2}\\
&\times e^{-D(Q^2+\kappa^2)|t|}.
\end{aligned}
\end{equation}
The factor $1/V$ uses the same intensive convention as the OCP spectrum. The charge mode relaxes with the Debye--Falkenhagen rate $\Omega_D(Q)=D(Q^2+\kappa^2)$ and Debye time $\tau_D=1/(D\kappa^2)$.

Using Eq.~\eqref{eq:W_from_charge}, the corresponding potential-energy spectrum is
\begin{equation}\label{eq:SDFT_potential_spectrum}
\begin{aligned}
\widetilde K_{\rm el}(Q,\omega)
&=\frac{8\pi C D\kappa^2}
{Q^2\left[\omega^2+D^2(Q^2+\kappa^2)^2\right]},\\
C&=\frac{k_{\mathrm B}Tq_0^2}{4\pi\varepsilon_b}.
\end{aligned}
\end{equation}
Its frequency integral gives
\begin{equation}\label{eq:SDFT_equal_time}
\int_{-\infty}^{\infty}\frac{d\omega}{2\pi}
\widetilde K_{\rm el}(Q,\omega)
=
4\pi C\frac{\kappa^2}{Q^2(Q^2+\kappa^2)},
\end{equation}
which is identical to the equal-time OCP spectrum in Eq.~\eqref{eq:static_spectrum_recovered}. The two models therefore differ only through the temporal organization of the same Gaussian static Coulomb fluctuations at this level.

\subsection{Exact on-shell line strength}

Substitution of Eq.~\eqref{eq:SDFT_potential_spectrum} into the general on-shell formula~\eqref{eq:Gdyn_spectral} gives
\begin{equation*}
\begin{aligned}
G_{\rm el}(v)
={}&\frac{2m^2C}{\pi\hbar^4}
\int_0^\infty\frac{dx}{x(1+x^2)}
\tan^{-1}\left(\frac{\nu x}{1+x^2}\right),\\
&\nu=\frac{v}{D\kappa}.
\end{aligned}
\end{equation*}
Denoting the dimensionless integral above by $I(\nu)$, differentiation gives
\begin{equation*}
\begin{aligned}
I'(\nu)
&=\int_0^\infty
\frac{dx}{(1+x^2)^2+\nu^2x^2}\\
&=\frac{\pi}{2\sqrt{\nu^2+4}}.
\end{aligned}
\end{equation*}
Since the integral vanishes at $\nu=0$, the exact result is
\begin{equation}\label{eq:Gdyn_el_exact}
G_{\rm el}(v)
=
\frac{m^2C}{\hbar^4}
\operatorname{arsinh}\left(
\frac{v}{2D\kappa}
\right).
\end{equation}
Equation~\eqref{eq:Gdyn_el_exact} is exact within the linearized SDFT model and the on-shell straight-line construction. In the weak-saddle branch,
\begin{equation*}
\ell_{\rm el}(k)=\frac{\hbar^4 k^2}{m^2 C\operatorname{arsinh}[v/(2D\kappa)]}.
\end{equation*}
For $v\ll D\kappa$,
\begin{equation*}
\begin{aligned}
G_{\rm el}(v)
&\simeq\frac{m^2C}{2\hbar^4}\frac{v}{D\kappa},\\
\ell_{\rm el}(k)
&\simeq\frac{2\hbar^2D\kappa v}{C}
=\frac{2\hbar^3D\kappa}{mC}\,k.
\end{aligned}
\end{equation*}
The formal slow expansion gives the same linear-in-$k$ weak-branch law as the slow collisionless OCP, albeit with a different microscopic velocity scale and prefactor. As shown in Sec.~\ref{sec:electrolyte_overlap}, this limit is not a physically accessible eikonal regime for conventional charged probes traversing ordinary liquid electrolytes. The experimentally relevant asymptote of the electrolyte model is instead the \emph{fast-probe} regime $v\gg D\kappa$, for which
\begin{equation}\label{eq:Gel_fast_dynamical_cutoff}
G_{\rm el}(v)\simeq\frac{m^2C}{\hbar^4}\ln\frac{v}{D\kappa}
=\frac{m^2C}{\hbar^4}\ln(\kappa L_D),
\qquad L_D=\frac{v}{D\kappa^2}.
\end{equation}
The Debye--Falkenhagen relaxation time $\tau_D=(D\kappa^2)^{-1}$ thus supplies a natural dynamical infrared length $L_D=v\tau_D$. This resolves, within the infinite-medium SDFT calculation, the infrared ambiguity of the quasi-static one-particle attenuation. When a macroscopic cutoff $L_{\rm IR}$ is present, the logarithm is controlled to leading accuracy by $L_{\rm eff}^{\rm el}=\min(L_{\rm IR},L_D)$. The intrinsic cutoff can matter only when $L_D$ is shorter than the relevant geometric or coherence scale. The high-velocity expression is an asymptote of the overdamped SDFT model, not a guarantee of microscopic validity at arbitrarily high frequencies. Inertial and solvent-relaxation channels may need to be included for real fast probes.

\subsection{Mutual coherence in the formal slow diffusive limit}

When $v\ll D\kappa$, the Doppler frequency $vQ_\parallel$ is small compared with $D(Q^2+\kappa^2)$ for all relevant wave numbers. Equation~\eqref{eq:SDFT_potential_spectrum} then reduces to
\begin{equation*}
\begin{aligned}
\widetilde K_{\rm el}(Q,vQ_\parallel)
={}&\frac{8\pi C\kappa^2}
{DQ^2(Q^2+\kappa^2)^2}\\
&\times\left[1+O\left(\frac{v^2}{D^2\kappa^2}\right)\right].
\end{aligned}
\end{equation*}
Substitution into Eq.~\eqref{eq:Dphi_thick_general} gives
\begin{equation}\label{eq:Dphi_el_scaling}
\mathcal D_d^{\rm el}(\rho)
\simeq
8\frac{d}{\ell_{\rm el}}\,
\mathcal F_{\rm el}(\kappa\rho),
\end{equation}
where
\begin{equation*}
\mathcal F_{\rm el}(x)
=
\frac{2}{\pi}
\int_0^\infty
\frac{dy}{(1+y^2)^2}
\left[
1-\frac{\sin(xy)}{xy}
\right].
\end{equation*}
The integral has the closed form
\begin{equation}\label{eq:JD_closed}
\mathcal F_{\rm el}(x)
=
\frac{1}{2}
-\frac{1-e^{-x}}{x}
+\frac{1}{2}e^{-x}.
\end{equation}
Its limiting forms are $\mathcal F_{\rm el}(x)=x^2/12+O(x^3)$ for $x\ll1$ and $\mathcal F_{\rm el}(x)=1/2-1/x+O(e^{-x})$ for $x\gg1$. Hence
\begin{equation}\label{eq:Dphi_el_small}
\mathcal D_d^{\rm el}(\rho)
\simeq
\frac{2}{3}
\frac{d}{\ell_{\rm el}}
(\kappa\rho)^2,
\qquad
\kappa\rho\ll1,
\end{equation}
while the large-separation plateau is
\begin{equation*}
\begin{aligned}
\mathcal D_d^{\rm el}(\infty)
&=\frac{4d}{\ell_{\rm el}},\\
\gamma_d^{\rm el}(\infty)
&=\exp\left(-\frac{2d}{\ell_{\rm el}}\right).
\end{aligned}
\end{equation*}
Since $\mathcal F_{\rm el}(x)$ increases monotonically from $0$ to $1/2$, a finite $e^{-1}$ coherence radius exists if and only if $d>\ell_{\rm el}/2$. If this radius lies in the small-separation regime, Eq.~\eqref{eq:Dphi_el_small} gives $\rho_c\simeq(\sqrt{3}/\kappa)\sqrt{\ell_{\rm el}/d}$. Unlike the collisionless OCP result in Eq.~\eqref{eq:rho_c_conditional}, the diffusive small-$\rho$ relation contains no logarithmic correction.

\section{Discussion}
\label{sec:discussion}

\subsection{Consistency with the quasi-static theory}

The dynamic and quasi-static problems share the same equal-time RPA correlator, Eq.~\eqref{eq:static_correlator_recovered}, but differ in the line integral sampled by a moving particle. In the quasi-static problem the $1/r$ tail makes the one-particle line strength infrared divergent, so the attenuation length contains the external scale $L_{\rm IR}$. In the dynamic problem the temporal spectrum restricts the Doppler-resonant modes. For fast particles this replaces $L_{\rm IR}$ by the smaller of $L_{\rm IR}$ and $v/\omega_{\rm p}$. For slow particles the line strength becomes finite and proportional to $v$.

The finite-thickness mutual-coherence observable is infrared finite in both theories. In the static theory the algebraic envelope occurs only in the intermediate window $\lambda_D\ll\rho\ll d$, followed by a finite-$d$ plateau~\cite{budkov2026static}. In the slow dynamic theory derived here, temporal decorrelation is stronger: already in the longitudinal thick-slab approximation the transverse structure function approaches the plateau $4d/\ell_{\rm slow}$. Thus the dynamic calculation reinforces, rather than circumvents, the conclusion that an $e^{-1}$ coherence radius is conditional.

\subsection{Collisionless OCP versus diffusive electrolyte}

The collisionless OCP and the diffusive electrolyte have the same Gaussian equal-time potential spectrum, Eqs.~\eqref{eq:static_spectrum_recovered} and \eqref{eq:SDFT_equal_time}, but their dynamical regularization mechanisms are different. In the OCP, long-wavelength fluctuations are organized by plasma oscillations and ballistic thermal motion, with $\omega_{\rm p}=\kappa v_{\rm th}$. In the electrolyte, the charge mode relaxes diffusively with the finite zero-wave-number rate $D\kappa^2$. The corresponding dynamical lengths sampled by a particle of velocity $v$ are $L_{\rm dyn}^{\rm OCP}\sim v/\omega_{\rm p}$ and $L_{\rm dyn}^{\rm el}=v/(D\kappa^2)$. Both lengths reproduce a Coulomb logarithm when $\kappa L_{\rm dyn}\gg1$, but the dimensionless velocity entering that logarithm is $v/v_{\rm th}$ for the collisionless plasma and $v/(D\kappa)$ for the electrolyte.

The slow limits also have a common \emph{formal} analytic structure:
\begin{equation*}
\begin{aligned}
G_{\rm OCP}(v)
&\simeq\frac{m^2C}{\hbar^4}
\frac{1}{\sqrt{2\pi}}\frac{v}{v_{\rm th}},\\
G_{\rm el}(v)
&\simeq\frac{m^2C}{\hbar^4}
\frac{1}{2}\frac{v}{D\kappa}.
\end{aligned}
\end{equation*}
Thus $G\propto v$ and $\ell\propto k$ in both mathematical limits, with coefficients probing thermal phase mixing and Debye charge relaxation, respectively. These asymptotes have different physical status: the slow OCP regime may overlap a controlled eikonal domain, while the slow-electrolyte limit does not do so for ordinary liquid electrolytes and experimentally usable coherent probes (Sec.~\ref{sec:electrolyte_overlap}). Conversely, the \emph{fast} diffusive-electrolyte limit is especially significant: the intrinsic length $L_D=v/(D\kappa^2)$ regularizes the Coulomb logarithm, unlike the externally cut off quasi-static model.

\begin{figure}[tb]
\centering
\includegraphics[width=0.82\linewidth]{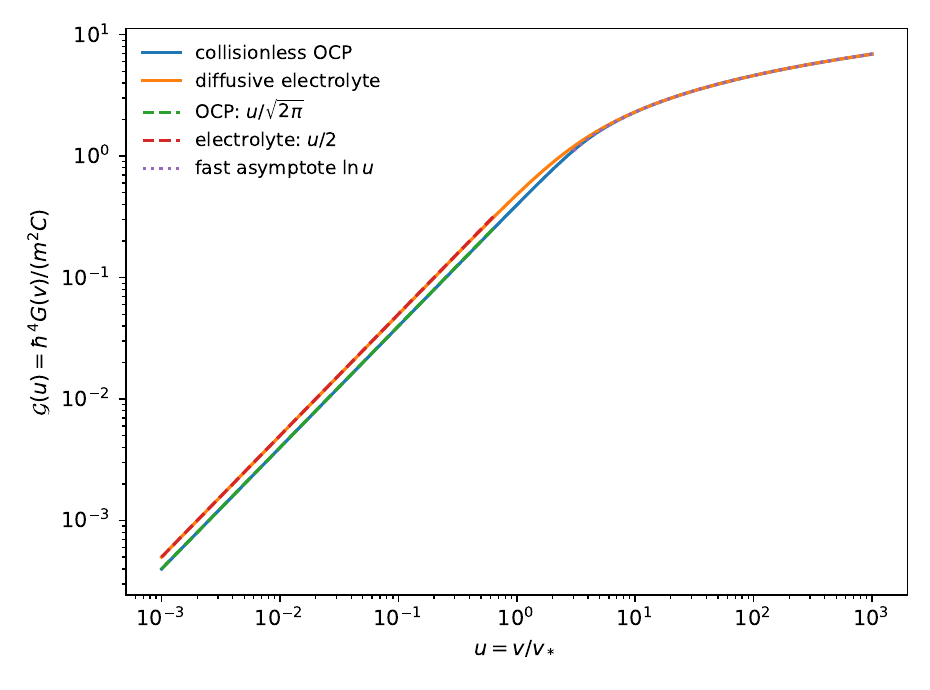}
\caption{Dimensionless dynamic line strength $\mathcal G(\zeta)=\hbar^4G(v)/(m^2C)$ as a function of the natural dimensionless velocity $\zeta=v/v_*$. For the collisionless OCP, $v_*=v_{\rm th}$ and the solid curve is obtained from Eq.~\eqref{eq:G_OCP_numeric}. For the diffusive electrolyte, $v_*=D\kappa$ and the solid curve is the exact SDFT result $\operatorname{arsinh}(\zeta/2)$. Dashed segments show the slow asymptotes $\zeta/\sqrt{2\pi}$ and $\zeta/2$, and the dotted segment shows the common fast asymptote $\ln\zeta$. The slow-electrolyte asymptote is a formal mathematical limit for ordinary liquid electrolytes. The experimentally relevant branch is the fast-probe side. The OCP high-velocity branch shown here assumes $v/\omega_{\rm p}<L_{\rm IR}$.}
\label{fig:attenuation_crossover}
\end{figure}

To display the complete crossover, it is convenient to introduce the dimensionless line strength $\mathcal G(\zeta)=\hbar^4G(v)/(m^2C)$ with $\zeta=v/v_*$, where $v_*=v_{\rm th}$ for the OCP and $v_*=D\kappa$ for the diffusive electrolyte. For the collisionless OCP, the $Q$ integration in Eq.~\eqref{eq:Gdyn_master} can be performed analytically, leaving the one-dimensional representation
\begin{equation}\label{eq:G_OCP_numeric}
\mathcal G_{\rm OCP}(\zeta)
=\frac{1}{\pi}\int_0^{\zeta/\sqrt{2}}
\frac{d\xi}{\xi}\,
\arg\left[1+\xi Z(\xi)\right],
\end{equation}
where the continuous branch of the argument is taken from $\arg[1+\xi Z(\xi)]=0$ at $\xi=0$. Figure~\ref{fig:attenuation_crossover} compares the numerical evaluation of Eq.~\eqref{eq:G_OCP_numeric} with the exact SDFT result in Eq.~\eqref{eq:Gdyn_el_exact}. For the OCP curve the plotted high-velocity branch corresponds to the dynamically limited case $v/\omega_{\rm p}<L_{\rm IR}$, so that the Coulomb logarithm is cut off by the plasma dynamics rather than by the external scale. At $\zeta=1$, the dimensionless line strengths are $\mathcal G_{\rm OCP}=0.396$ and $\mathcal G_{\rm el}=0.481$. The OCP slow asymptote differs from the full result by less than $1\%$ at $\zeta=1$, whereas the SDFT result differs from $\zeta/2$ by about $4\%$. On the fast side, the OCP curve is within $1\%$ of $\ln\zeta$ at $\zeta=3$. The diffusive result approaches the logarithm more gradually and is within $0.5\%$ by $\zeta=10$. Thus both models interpolate between the common linear law $G\propto v$ and a Coulomb-logarithmic regime, but the crossover shape retains information about the microscopic relaxation mechanism.

The difference is especially clear in the transverse coherence crossover. Writing $\mathcal F_{\rm OCP}(x)\equiv\mathcal F(x)$ for the collisionless result of Eq.~\eqref{eq:J_integral}, the small-$x$ forms are $\mathcal F_{\rm OCP}(x)\simeq(x^2/6)\ln(1/x)$ and $\mathcal F_{\rm el}(x)\simeq x^2/12$, where only the leading singular structure of Eq.~\eqref{eq:J_small} is displayed for the OCP. At large $x$,
\begin{equation}\label{eq:J_large_compare}
\begin{aligned}
\mathcal F_{\rm OCP}(x)
&=\frac{1}{2}-\frac{1}{x^2}+O(x^{-4}),\\
\mathcal F_{\rm el}(x)
&=\frac{1}{2}-\frac{1}{x}+O(e^{-x}).
\end{aligned}
\end{equation}
Both monotone slow-regime models have the thick-slab plateau $\mathcal D_d(\infty)\simeq4d/\ell$ and the same threshold for a finite $e^{-1}$ radius. Their approach to the plateau differs. The collisionless response is logarithmically nonanalytic at small transverse separation and approaches saturation as $1/\rho^2$. The diffusive response has no small-$\rho$ logarithm and approaches saturation more slowly, as $1/\rho$. Thus the equal-time correlator alone does not fix the spatial dephasing law. The temporal relaxation spectrum remains observable in the mutual coherence.

\begin{figure}[tb]
\centering
\includegraphics[width=0.82\linewidth]{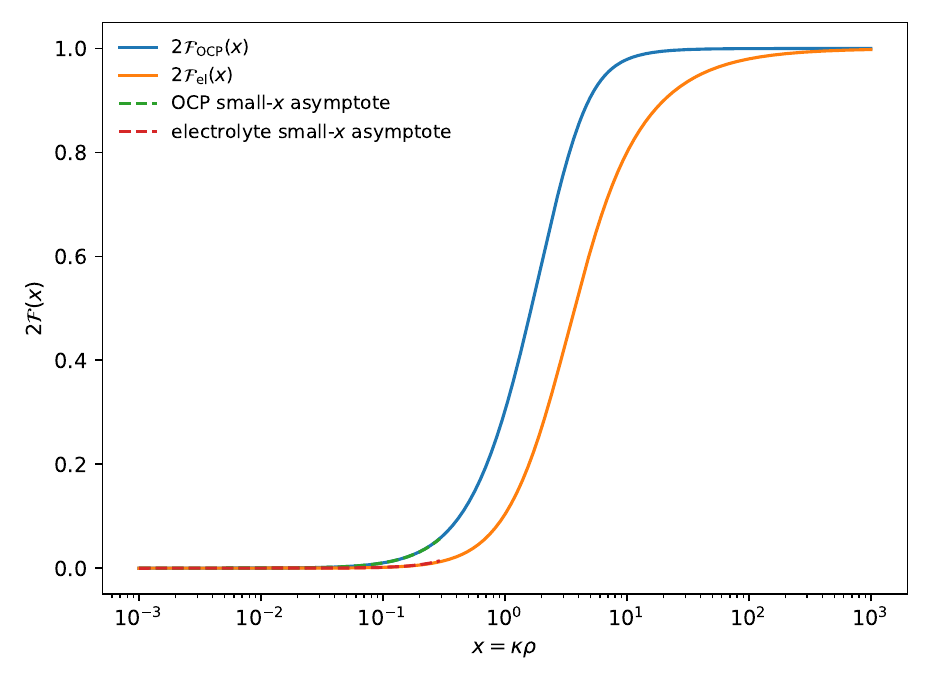}
\caption{Normalized slow-regime transverse scaling functions for the collisionless OCP and the diffusive electrolyte. The electrolyte branch illustrates a formal slow-diffusion limit, not an experimentally accessible eikonal regime in ordinary liquid electrolytes. Solid curves are the exact functions $2\mathcal F_{\rm OCP}(x)$ and $2\mathcal F_{\rm el}(x)$, with $x=\kappa\rho$. Dashed segments show their small-$x$ asymptotes. Both curves approach the same plateau value $1$, but the crossover occurs on different transverse scales.}
\label{fig:coherence_scaling}
\end{figure}

The difference is visible over the entire mathematical crossover, not only in the asymptotic expansions. The slow-electrolyte curves below are theoretical benchmarks rather than predictions for conventional liquid-electrolyte probes. Figure~\ref{fig:coherence_scaling} shows the normalized functions $2\mathcal F_{\rm OCP}(x)$ and $2\mathcal F_{\rm el}(x)$, both of which approach unity at large $x$. At $x=1$, the two values are $0.303$ and $0.104$, respectively. At $x=10$, they are $0.979$ and $0.800$. Hence, at the same value of $\kappa\rho$ and the same normalized plateau amplitude, the collisionless OCP reaches the saturation regime substantially earlier than the diffusive electrolyte.

\begin{figure}[tb]
\centering
\includegraphics[width=0.82\linewidth]{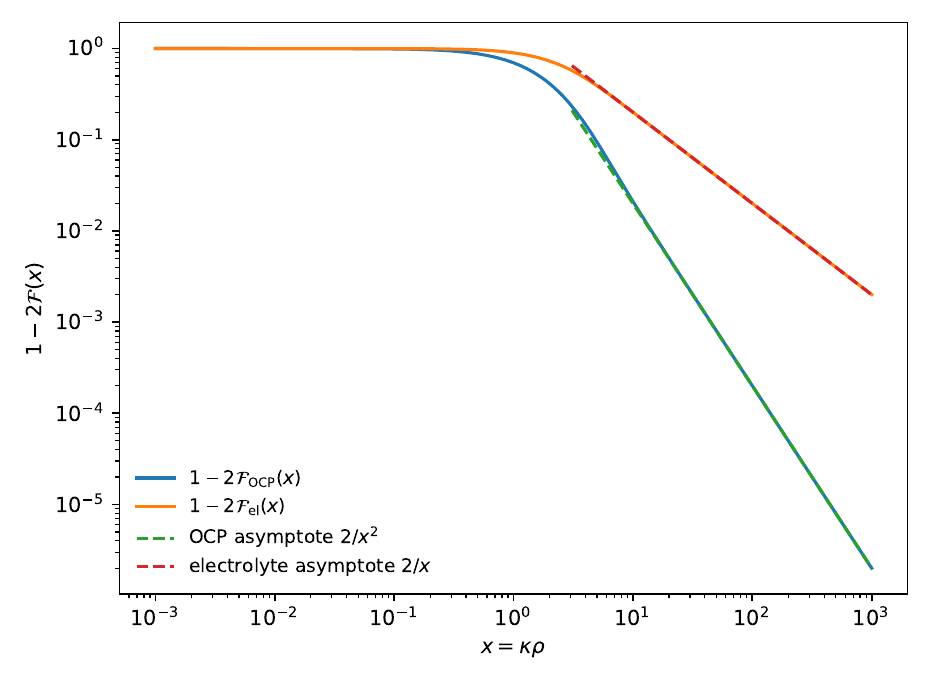}
\caption{Approach of the slow-regime coherence scaling functions to their common plateau. The electrolyte branch illustrates a formal slow-diffusion limit, not an experimentally accessible eikonal regime in ordinary liquid electrolytes. The plotted deficit is $1-2\mathcal F(x)$. Dashed lines show the asymptotic laws $2/x^2$ for the collisionless OCP and $2/x$ for the diffusive electrolyte.}
\label{fig:plateau_approach}
\end{figure}

The distinct large-$x$ tails are more transparent when the distance from the plateau is plotted directly. Figure~\ref{fig:plateau_approach} shows $1-2\mathcal F(x)$ on logarithmic axes. The exact curves approach the predicted power laws $2/x^2$ for the collisionless OCP and $2/x$ for the diffusive electrolyte. For example, at $x=10$ the plateau deficits are $2.10\times10^{-2}$ and $2.00\times10^{-1}$, respectively. The order-of-magnitude separation at this moderate value of $x$ illustrates that the temporal relaxation law can strongly affect the finite-separation crossover even though the equal-time correlators coincide.

\begin{figure}[tb]
\centering
\includegraphics[width=0.82\linewidth]{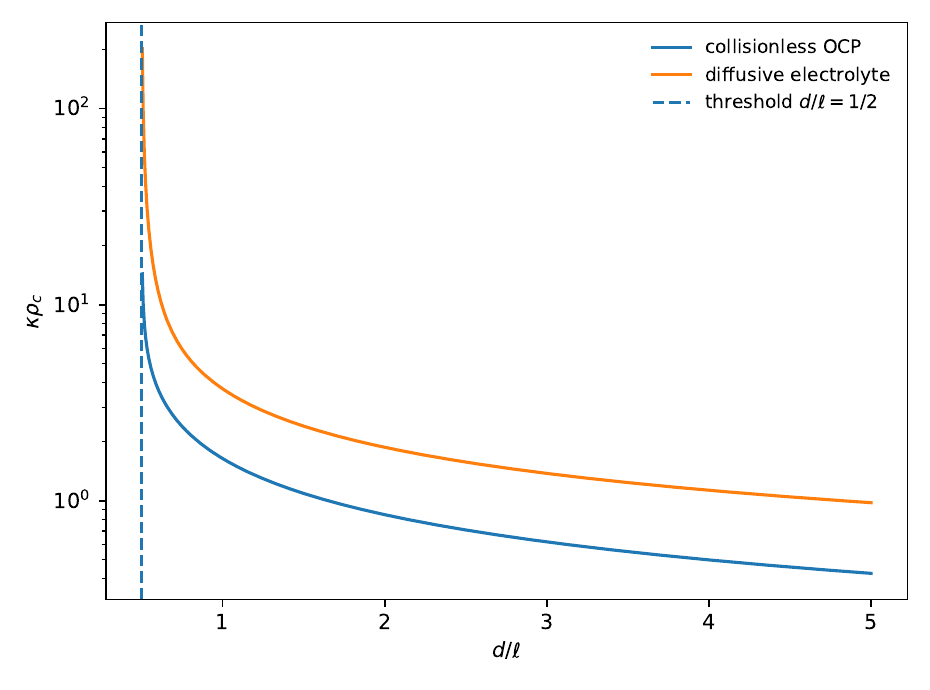}
\caption{Dimensionless coherence radius $\kappa\rho_c$ as a function of $d/\ell$ in the two slow-regime models. The vertical dashed line marks the common existence threshold $d/\ell=1/2$. The attenuation length $\ell$ is the corresponding weak-saddle value for each model. The electrolyte slow limit is a formal benchmark for ordinary liquid electrolytes.}
\label{fig:coherence_radius}
\end{figure}

The same distinction controls the coherence radius close to its existence threshold. Let $x_c=\kappa\rho_c$, with $\ell$ denoting the corresponding slow weak-saddle attenuation length. Since $\mathcal D_d(\rho)=8(d/\ell)\mathcal F(\kappa\rho)$ in both models, the finite $e^{-1}$ radius is determined by
\begin{equation}\label{eq:coherence_radius_numeric}
\mathcal F(x_c)=\frac{\ell}{4d},
\qquad \frac{d}{\ell}>\frac{1}{2}.
\end{equation}
Numerical solutions of Eq.~\eqref{eq:coherence_radius_numeric} are shown in Fig.~\ref{fig:coherence_radius}. The diffusive electrolyte has the larger transverse coherence radius throughout the plotted range. At $d/\ell=0.6$, one finds $x_c=3.74$ for the OCP and $x_c=12.0$ for the electrolyte. At $d/\ell=1$, the values are $1.64$ and $3.72$, respectively. At $d/\ell=2$, they are $0.849$ and $1.87$.

\begin{figure}[tb]
\centering
\includegraphics[width=0.82\linewidth]{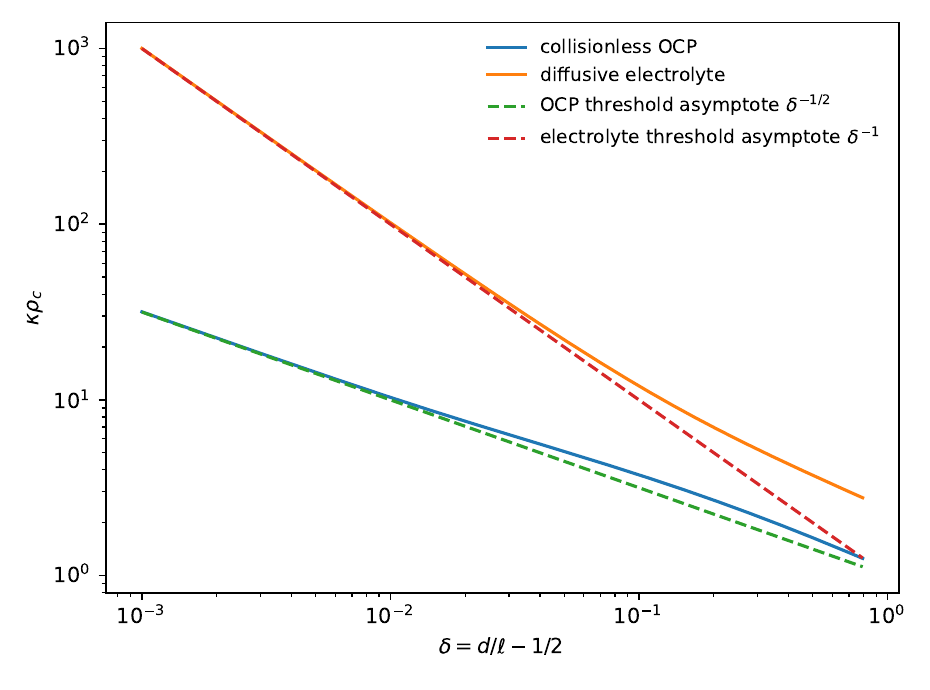}
\caption{Threshold divergence of the dimensionless coherence radius, with $\delta=d/\ell-1/2$. The electrolyte branch illustrates a formal slow-diffusion limit, not an experimentally accessible eikonal regime in ordinary liquid electrolytes. Solid curves are numerical solutions of Eq.~\eqref{eq:coherence_radius_numeric}. Dashed lines show the asymptotes $\kappa\rho_c\sim\delta^{-1/2}$ for the collisionless OCP and $\kappa\rho_c\sim\delta^{-1}$ for the diffusive electrolyte.}
\label{fig:threshold_scaling}
\end{figure}

The divergence at the threshold is itself model dependent. Writing $\delta=d/\ell-1/2$ with $\delta\to0^+$, Eq.~\eqref{eq:coherence_radius_numeric} together with Eq.~\eqref{eq:J_large_compare} gives
\begin{equation}\label{eq:coherence_threshold_exponents}
\begin{aligned}
\kappa\rho_c\simeq
\begin{cases}
\delta^{-1/2}, & \text{collisionless OCP},\\[3pt]
\delta^{-1}, & \text{diffusive electrolyte}.
\end{cases}
\end{aligned}
\end{equation}
Figure~\ref{fig:threshold_scaling} verifies these two threshold exponents using the full scaling functions. A log--log fit over $\delta<10^{-2}$ gives slopes $-0.488$ for the OCP and $-0.993$ for the electrolyte, approaching the analytic values $-1/2$ and $-1$. Thus the common criterion $d/\ell>1/2$ does not imply a common threshold law: the divergence of the transverse coherence radius provides a sharp theoretical fingerprint of the dynamical class, although the electrolyte slow-limit exponent is not directly measurable in an ordinary liquid-cell beam experiment.

\subsection{Validity of the saddle results}

The slow weak-saddle law $\ell_{\rm slow}(k)\propto k$ follows under the simultaneous assumptions $v\ll v_{\rm th}$, $g(v)\ll1$, and a valid straight-line approximation. The closed mutual-coherence function in Eq.~\eqref{eq:Dphi_slow_scaling} applies in the same on-shell slow regime, with the additional thick-slab condition.

The saturation $\ell\to g_0^{-1/2}$ is less direct. It follows from analytically continuing the leading slow-response action to a complex saddle with $g_0/k^2\gg1$. It is therefore best regarded as a formal prediction of the self-consistent eikonal saddle, not as a rigorously controlled low-energy asymptote. A full calculation beyond the straight-line approximation would be needed to establish whether the saturation survives transverse path fluctuations.

\subsection{Physical accessibility of the plasma and electrolyte regimes}
\label{sec:electrolyte_overlap}

The analytical slow limits of the collisionless OCP and diffusive
 electrolyte should not be assigned the same experimental status. For a
 nonrelativistic coherent probe, the spatial eikonal condition is
 $k\gg\kappa$, while $v=\hbar k/m$. The slow \emph{electrolyte} condition
 $v\ll D\kappa$ would therefore require
\begin{equation}\label{eq:slow_eikonal_overlap}
\kappa\ll k\ll\frac{mD\kappa}{\hbar},
\qquad
\frac{\hbar\kappa}{m}\ll v\ll D\kappa.
\end{equation}
A parametrically broad overlap demands
\begin{equation}\label{eq:slow_eikonal_mass_condition}
\mathcal R_{\rm el}\equiv\frac{mD}{\hbar}\gg1.
\end{equation}
The criterion is independent of $\kappa$, so changing salt concentration
 alone cannot remedy a missing overlap. With
 $D\sim10^{-9}\,\mathrm{m^2\,s^{-1}}$ and
 $\hbar/m_e\simeq1.16\times10^{-4}\,\mathrm{m^2\,s^{-1}}$,
 one finds $\mathcal R_{\rm el}\sim8.6\times10^{-6}$ for an electron.
 Even for a proton or an alpha particle the corresponding ratios are
 only about $1.6\times10^{-2}$ and $6.3\times10^{-2}$. Although very
 massive ions can improve the kinematic ratio, attaining a coherent,
 essentially ballistic, ultra-slow trajectory through a liquid introduces
 severe stopping, scattering, and charge-exchange difficulties that are
 outside the present phase-noise model. Thus the slow diffusive
 electrolyte regime should presently be regarded as \emph{experimentally
 inaccessible in ordinary liquid electrolytes and formal within the
 eikonal theory}, not as a prediction for available electron or ion
 beam measurements. This is a statement about the physically relevant
 domain, not a mathematical nonexistence theorem for every conceivable
 medium and probe.

In fact, the same estimates identify the \emph{fast} electrolyte regime
 as the natural eikonal limit. Its dimensionless velocity is
\begin{equation}\label{eq:fast_el_from_eikonal}
\frac{v}{D\kappa}
=\frac{k/\kappa}{\mathcal R_{\rm el}}\gg1
\qquad (k\gg\kappa,\ \mathcal R_{\rm el}\ll1).
\end{equation}
Accordingly, the central experimentally oriented electrolyte result is
 not the formal slow-limit law $\ell_{\rm el}\propto k$ or its associated
 slow transverse scaling function. It is the exact SDFT crossover and,
 in the fast-probe sector, the dynamical Coulomb logarithm in
 Eq.~\eqref{eq:Gel_fast_dynamical_cutoff}. The diffusive charge-relaxation
 time generates the natural length
 $L_D=v/(D\kappa^2)$, so the infinite-medium line strength is finite
 without an ad hoc infrared cutoff, unlike the quasi-static theory.
 A finite slab or other macroscopic scale may nonetheless set a smaller
 effective length, $L_{\rm eff}^{\rm el}=\min(L_{\rm IR},L_D)$, and the
 overdamped description remains conditional on the frequencies sampled
 by the probe. Therefore, intrinsic dynamical regularization is a
 robust \emph{theoretical distinction}, but its direct experimental
 isolation requires a suitable time-of-flight and length-scale window.

For the collisionless \emph{plasma}, by contrast, there is no analogous
 small ion-diffusion parameter constraining the probe to the fast limit.
 The slow-plasma overlap conditions are
\begin{equation}\label{eq:slow_plasma_overlap}
\frac{\hbar\kappa}{m}\ll v\ll v_{\rm th},
\qquad
\frac{m v_{\rm th}}{\hbar\kappa}\gg1,
\end{equation}
which can in principle be realized for appropriate plasma densities,
 temperatures, and sufficiently massive coherent probes. Faster probes,
 $v\gg v_{\rm th}$, access the complementary Coulomb-logarithmic branch
 with $L_{\rm dyn}^{\rm OCP}\sim v/\omega_{\rm p}$. Both the slow and the
 fast OCP regimes therefore merit independent physical consideration,
 subject in each case to weak-saddle control, paraxial coherence, and
 suppression or characterization of ordinary collisions. This contrasts
 with the ordinary liquid electrolyte, where the slow branch serves
 primarily as a mathematically useful dynamical-class comparison.

\subsection{Perspectives for coherent-probe experiments}
\label{sec:experimental_outlook}

The present OCP results suggest an experimental diagnostic based on the
velocity dependence of two-path interference contrast, provided that the
incoming particle remains spatially coherent and that the eikonal,
weak-saddle, and relevant finite-thickness conditions are met. In the
controlled slow-OCP and thick-slab regime,
Eq.~\eqref{eq:Dphi_slow_scaling} predicts, at fixed $\rho$, $d$, and
plasma parameters,
\begin{equation}\label{eq:slow_plasma_visibility_velocity}
-\ln\!\frac{\mathcal V(\rho,v)}{\mathcal V_0(\rho,v)}
=\frac{4dC}{\sqrt{2\pi}\,\hbar^2v v_{\rm th}}
\mathcal F_{\rm OCP}(\kappa\rho)
\propto v^{-1}.
\end{equation}
This differs from the $v^{-2}$ dependence of the phase-structure
function in the frozen finite-slab limit. These are predictions of the
\emph{one-component, Gaussian, rectilinear} model in their respective
validity windows, not universal visibility laws for an arbitrary plasma.
The finite-separation contrast is also less sensitive than the mean
one-particle amplitude to common long-wavelength phase fluctuations.
Therefore, the intrinsic dynamical infrared cutoff cannot in general
be inferred from a single two-ray visibility measurement.

A coherent charged-particle probe, possibly prepared with ion-beam
optics, could in principle test this velocity dependence. A quantitative
proposal for an ordinary quasineutral electron--ion plasma, however,
requires the full multicomponent dynamical dielectric response together
with microscopic collision, charge-exchange, and non-Gaussian effects.
These extensions lie beyond the present study. For diffusive liquid
electrolytes, the relevant theoretical prediction remains the fast-probe
dynamical regularization with $L_D=v/(D\kappa^2)$, whereas a finite
liquid cell can be too thin for this bulk infrared scale to develop.
Experimental realization of either case requires a separate feasibility
analysis, including the initial coherence of the probe and competing
scattering processes. No numerical interferometric contrast for a
specific two-component plasma or liquid-cell geometry is claimed here.

\subsection{Experimental access to the attenuation length}

Because $\ell$ is an amplitude attenuation length rather than a transport mean free path, its most direct operational signature within the present phase-noise model is a coherence observable. In principle, the weak on-shell $\ell$ can be inferred from a diffraction measurement of the transverse mutual coherence rather than from a direct measurement of the disorder-averaged one-particle propagator. Consider a two-beam diffraction geometry in which two paraxial components traverse the same ionic layer while sampling transversely separated paths. If $I_1$ and $I_2$ are the intensities of the two components measured separately, the fringe visibility is
\begin{equation}
\mathcal V(\rho)=
\frac{2\sqrt{I_1I_2}}{I_1+I_2}\,
|\gamma_d(\rho)|.
\label{eq:visibility_gamma}
\end{equation}
Thus, after normalization by a reference measurement without the fluctuating ionic contribution, the ratio $\mathcal V(\rho)/\mathcal V_0(\rho)$ directly gives $|\gamma_d(\rho)|$, provided that the instrumental coherence and the relative beam intensities are unchanged. For balanced diffraction channels the prefactor in Eq.~\eqref{eq:visibility_gamma} is unity.

In the dynamic thick-slab weak branch, Eq.~\eqref{eq:plateau_ell_relation} gives $\gamma_d(\infty)=\exp(-2d/\ell)$. If a measurable large-separation plateau can be reached within the controlled thick-slab regime, the attenuation length can in principle be inferred from the diffraction contrast,
\begin{equation}
\ell=
-\frac{2d}{\ln\!\left[\mathcal V(\infty)/\mathcal V_0(\infty)\right]}.
\label{eq:ell_from_visibility}
\end{equation}
When the plateau remains resolvable, an alternative is to repeat the measurement for several layer thicknesses. Then
\begin{equation}
-\ln\!\left[\frac{\mathcal V(\infty,d)}{\mathcal V_0(\infty,d)}\right]
=\frac{2d}{\ell},
\label{eq:visibility_thickness_slope}
\end{equation}
so that the slope of the measured logarithmic visibility with respect to $d$ is $2/\ell$. Alternatively, when the controlled slow-plasma conditions are satisfied, the full transverse diffraction envelope can be fitted to $\ln[\mathcal V(\rho)/\mathcal V_0(\rho)]=-4(d/\ell)\mathcal F_{\rm OCP}(\kappa\rho)$. The corresponding slow-electrolyte function is a formal theoretical comparison and should not be used as an electron-beam fitting formula in ordinary liquid electrolytes (Sec.~\ref{sec:electrolyte_overlap}). Fast-electrolyte measurements, if feasible after controlling competing scattering and solvent channels, must instead be analyzed with the finite-thickness spectral expression, Eq.~\eqref{eq:Dphi_finite_spectral}, evaluated using the full diffusive spectrum, rather than with the slow-limit transverse scaling function.

The factor $2d/\ell$ in the large-separation plateau has a simple interpretation: for two statistically independent distant rays, the normalized mutual coherence contains the product of two one-particle coherent-amplitude factors. This operational relation is specific to the integrable dynamic line correlator and the thick-slab regime in which Eq.~\eqref{eq:plateau_ell_relation} is valid. It should not be transferred unchanged to the quasi-static finite-slab problem, where the mutual-coherence plateau contains the finite-thickness function $H(\kappa d)$ while the one-particle attenuation scale remains sensitive to the infrared regularization.

In an electron-beam realization, ordinary elastic angular scattering and electronic inelastic energy-loss channels are additional processes that are not contained in the ionic fluctuation spectrum used here~\cite{Egerton2011}. They can reduce or redirect the experimentally accessible paraxial component without being identical to the Coulomb phase-decorrelation mechanism calculated in this work. Any extraction of $\ell$ from diffraction contrast therefore requires these external channels to be controlled or independently characterized.

\subsection{Physical interpretation and extensions}

Temporal decorrelation weakens the accumulation of a common random phase from long-wavelength Coulomb modes. In the weak branch the dynamic attenuation rate grows as $1/k$ for slow particles, compared with the quasi-static $1/k^2$ dependence at fixed infrared cutoff. The dynamic environment therefore suppresses the frozen-disorder attenuation, although it does not eliminate attenuation as $k$ is lowered.

The collisionless OCP and diffusive symmetric electrolyte provide two analytically tractable dynamical classes with distinct experimental emphases. For plasmas, both the slow and fast branches may be accessible. For ordinary liquid electrolytes, the fast-probe branch and its natural dynamical infrared scale are the principal physically relevant results. Ionic liquids and concentrated electrolytes may additionally display unequal ionic diffusivities, viscoelastic relaxation, nonlocal dielectric response, and coupling to explicit solvent polarization. Equations~\eqref{eq:Gdyn_spectral}, \eqref{eq:Dphi_finite_spectral}, and \eqref{eq:Dphi_thick_general} remain the appropriate general formulas once the corresponding positive dynamic potential spectrum is supplied.

The finite-frequency spectra used above describe the dynamics of the ionic charge degrees of freedom. They do not constitute a complete electronic energy-loss function of a molecular liquid and therefore do not include solvent electronic excitations, ionization, or other microscopic inelastic channels. For relativistic electron beams the paraxial phase coefficient is $a_{\rm rel}=1/(\hbar v)$, as derived in Ref.~\cite{budkov2026static}. The finite-thickness coherence formulas retain their form under $a\to a_{\rm rel}$. The slow regimes require $v\ll v_{\rm th}$ for the OCP or $v\ll D\kappa$ for the electrolyte and are intrinsically nonrelativistic. Conventional transmission-electron-microscopy beams lie deep in the corresponding fast regimes. The slow-particle crossover is therefore not a direct prediction for sub-keV liquid-cell TEM.

\section{Conclusions}

We have developed a dynamic extension of the quasi-static Coulomb-fluctuation theory of Ref.~\cite{budkov2026static}. The one-particle scale is the eikonal attenuation length of the disorder-averaged propagator. In the controlled weak on-shell regime, it is the spatial counterpart of the single-particle spectral linewidth, Eq.~\eqref{eq:ell_spectral_meaning}. It is neither a transport mean free path nor an Anderson localization length. Its dimensionless saddle parameter is $g(v)=G(v)/k^3$, while the straight-line condition remains an independent requirement.

For the collisionless OCP, the classical fluctuation--dissipation theorem and the RPA longitudinal dielectric function give a dynamic spectrum whose frequency integral reproduces the quasi-static equal-time correlator. Fast probes recover the Coulomb logarithm with $L_{\rm eff}=\min(L_{\rm IR},v/\omega_{\rm p})$, whereas slow probes obey $G\propto v$ and hence $\ell\propto k$ in the controlled weak branch. Unlike the liquid-electrolyte case, both plasma regimes can possess a physically accessible eikonal overlap for suitable probe masses and plasma parameters.

The saddle-dependent velocity $v_s=v/s$ produces a formal slow-response interpolation, Eq.~\eqref{eq:slow_saddle_rate}, whose strong-saddle limit saturates at $\ell\sim g_0^{-1/2}$. This result is explicitly separated from the conditions required for analytic continuation and straight-line validity.

For mutual coherence, the exact finite-thickness expression is Eq.~\eqref{eq:Dphi_finite_spectral}. In the slow thick-slab weak branch, the complete transverse dependence is described by the scaling function $\mathcal F(\kappa\rho)$. The structure function saturates at $4d/\ell_{\rm slow}$, and an $e^{-1}$ coherence radius exists only when this plateau exceeds two. The estimate $\rho_c\sim\lambda_D\sqrt{\ell_{\rm slow}/d}$ is therefore a conditional small-separation relation, not a universal law. This conclusion is consistent with the finite-slab interpretation of the quasi-static problem.

For a symmetric diffusive electrolyte, linearized SDFT yields the exact on-shell result $G_{\rm el}(v)=(m^2C/\hbar^4)\operatorname{arsinh}[v/(2D\kappa)]$. The principal physically relevant regime in ordinary liquid electrolytes is $v\gg D\kappa$: the logarithmic fast-probe asymptote contains the intrinsic infrared length $L_D=v/(D\kappa^2)$, which removes the arbitrary infinite-medium cutoff of the quasi-static attenuation theory. A finite sample can impose the smaller effective length $\min(L_{\rm IR},L_D)$, and microscopic high-frequency processes delimit the range of validity of the overdamped spectrum. By contrast, the electrolyte's slow-$v$ law $\ell\propto k$ and its closed transverse-coherence scaling function are formal asymptotes: the simultaneous conditions $v\ll D\kappa$ and $k\gg\kappa$ demand $mD/\hbar\gg1$, which is severely violated for electrons and remains unfavorable for light ions in ordinary liquids. Consequently, the distinct slow-electrolyte plateau tails and threshold exponents derived here are mathematical fingerprints of diffusive charge relaxation, rather than predictions for an experimentally accessible slow-electrolyte beam regime. The slow-plasma results, on the other hand, can retain direct physical significance. This distinction is central to interpreting identical equal-time Coulomb correlations in media with different temporal dynamics.

The experimental outlook is most concrete for the collisionless OCP: within the controlled slow thick-slab domain the logarithmic interference visibility varies as $v^{-1}$, in contrast to the $v^{-2}$ frozen finite-slab limit. This distinction follows from the present analytical model and offers a possible diagnostic of dynamical Coulomb fluctuations. Whether it can be isolated in a real plasma requires separate work on coherent beam preparation, multicomponent response and microscopic scattering. For liquid electrolytes, the intrinsic dynamic infrared scale of the fast-probe theory is a robust bulk result, but its manifestation in a finite experimental cell must not be assumed without checking the transit and relaxation scales. Quantitative design of ion-beam interferometry and a theory of non-Gaussian, short-range collisions are reserved for future investigations.

\appendix

\section{Gaussian transverse fluctuations about the rectilinear path}
\label{app:dynamic_transverse_fluctuations}

The rectilinear-path approximation used above neglects fluctuations of the trajectory about the straight ray. Here we estimate the leading Gaussian contribution of the two transverse components in the dynamic problem. Starting from Eq.~\eqref{eq:averaged_dynamic_G}, write $\boldsymbol\eta(u)=\eta_{\parallel}(u)\mathbf n+\boldsymbol\eta_\perp(u)$ with $\mathbf n\cdot\boldsymbol\eta_\perp(u)=0$, and set $\eta_{\parallel}=0$. For $\Delta u=|u-u'|$ and $\Delta\boldsymbol\eta_\perp=\boldsymbol\eta_\perp(u)-\boldsymbol\eta_\perp(u')$, the temporal separation remains $\Delta u/v_s$, whereas the spatial separation becomes $R=[(\Delta u)^2+(s/k)|\Delta\boldsymbol\eta_\perp|^2]^{1/2}$. We therefore define
\begin{equation*}
J_{v_s}(\Delta u)
=-\left.
\frac{1}{R}\frac{\partial K(R,\Delta u/v_s)}{\partial R}
\right|_{R=\Delta u},
\qquad v_s=\frac{v}{s}.
\end{equation*}
The linear transverse variation vanishes by symmetry, and the quadratic part of the exponent relative to the straight path is
\begin{equation*}
\begin{aligned}
\mathcal Q_{\perp}^{(2)}
={}&\frac{i}{2}\int_0^rdu\,
\dot{\boldsymbol\eta}_{\perp}^{\,2}(u)\\
&+\frac{m^2s^3}{4\hbar^4k^3}
\int_0^rdu\int_0^rdu'\,
J_{v_s}(|u-u'|)\\
&\hspace{2.5em}\times
\left|\boldsymbol\eta_\perp(u)-\boldsymbol\eta_\perp(u')\right|^2.
\end{aligned}
\end{equation*}
For $r\to\infty$, let $\lambda$ denote the one-dimensional wave number conjugate to $u$, so that $\boldsymbol\eta_\perp(u)=\int(d\lambda/2\pi)e^{i\lambda u}\boldsymbol\eta_\perp(\lambda)$. The finite combination entering the quadratic kernel is
\begin{equation*}
\Delta J_v(\lambda)=2\int_0^\infty d\Delta u\,
J_v(\Delta u)\left[1-\cos(\lambda\Delta u)\right].
\end{equation*}
The quadratic exponent then has the same Gaussian form as in the quasi-static problem,
\begin{equation*}
\mathcal Q_{\perp}^{(2)}
=\frac12\int\frac{d\lambda}{2\pi}
\left[
 i\lambda^2+
 \frac{m^2s^3}{\hbar^4k^3}\Delta J_{v_s}(\lambda)
\right]
\left|\boldsymbol\eta_\perp(\lambda)\right|^2.
\end{equation*}
Defining $X_{\lambda}=(m^2s^3/\hbar^4k^3)\Delta J_{v_s}(\lambda)/\lambda^2$, normalization by the free Fresnel determinant produces the factor $1-iX_{\lambda}$. Since $\ln(1-iX)=\frac12\ln(1+X^2)-i\arctan X$, the term linear in $X$ is purely phase-like, whereas the first attenuation correction is quadratic. Including both transverse components gives
\begin{equation}\label{eq:deltaGamma_transverse_dynamic}
\delta\Gamma_\perp(s)
=
\int_0^\infty\frac{d\lambda}{2\pi}
\ln\left[
1+
\left(
\frac{m^2s^3}{\hbar^4k^3}
\frac{\Delta J_{v_s}(\lambda)}{\lambda^2}
\right)^2
\right].
\end{equation}
The associated displacement of the longitudinal saddle is of higher order on the weak branch, so below we set $s=1$.

A useful representation follows directly from the dynamic potential spectrum. Choosing the propagation direction as the $z$ axis and one transverse direction as $x$,
\begin{equation}\label{eq:DeltaJ_dynamic_spectral}
\begin{aligned}
\widehat J_v(\lambda)
&=v\int\frac{d^3Q}{(2\pi)^3}\,
Q_x^2\,
\widetilde K\!\left(Q,v(Q_z-\lambda)\right),\\
\Delta J_v(\lambda)&=\widehat J_v(0)-\widehat J_v(\lambda).
\end{aligned}
\end{equation}
The difference in Eq.~\eqref{eq:DeltaJ_dynamic_spectral}, rather than either term separately, is the quantity entering the transverse determinant.

\subsection{Diffusive electrolyte}

For the SDFT spectrum in Eq.~\eqref{eq:SDFT_potential_spectrum}, introduce $p=\lambda/\kappa$ and $\nu=v/(D\kappa)$. Equation~\eqref{eq:DeltaJ_dynamic_spectral} gives
\begin{equation*}
\Delta J_{\rm el}(\lambda)
=C\kappa^2\Delta j_{\rm el}(p,\nu),
\end{equation*}
where $D_0=(1+x^2)^2+\nu^2x^2\mu^2$ and $D_p=(1+x^2)^2+\nu^2(x\mu-p)^2$, with
\begin{equation*}
\Delta j_{\rm el}(p,\nu)
=\frac{\nu}{\pi}\int_0^\infty dx\,x^2
\int_{-1}^{1}d\mu\,(1-\mu^2)
\left(\frac{1}{D_0}-\frac{1}{D_p}\right).
\end{equation*}
In the slow regime $\nu\ll1$, the nontrivial longitudinal crossover occurs for $y=\nu p=O(1)$. At fixed $y$,
\begin{equation*}
\Delta j_{\rm el}(p,\nu)
=\nu H_{\rm el}(y)+O(\nu^3),
\end{equation*}
where
\begin{equation*}
\begin{aligned}
H_{\rm el}(y)
&=\frac{4}{3\pi}\int_0^\infty dx\,x^2
\left[
\frac{1}{(1+x^2)^2}
-
\frac{1}{(1+x^2)^2+y^2}
\right]\\
&=\frac13\left[
1-\sqrt{\frac{2}{1+\sqrt{1+y^2}}}
\right].
\end{aligned}
\end{equation*}
In particular, $H_{\rm el}(y)=y^2/24+O(y^4)$ for $y\to0$, while $H_{\rm el}(y)\to1/3$ for $y\to\infty$. Thus, with the same dimensionless local coupling as in the quasi-static calculation, $\chi=m^2C/(\hbar^4k^3)$, Eq.~\eqref{eq:deltaGamma_transverse_dynamic} becomes
\begin{equation*}
\delta\Gamma_\perp^{\rm el}
=\frac{\kappa}{2\pi\nu}
\int_0^\infty dy\,
\ln\left[
1+\chi^2\nu^6
\left(\frac{H_{\rm el}(y)}{y^2}\right)^2
\right].
\end{equation*}
Its leading algebraic term is
\begin{equation}\label{eq:deltaGamma_transverse_el}
\begin{aligned}
\delta\Gamma_\perp^{\rm el}
&\simeq
\frac{\kappa\chi^2\nu^5}{2\pi}I_\perp^{\rm el}
=3.75424\times10^{-4}\,\kappa\chi^2\nu^5,\\
I_\perp^{\rm el}
&=\int_0^\infty dy\,
\left(\frac{H_{\rm el}(y)}{y^2}\right)^2
=\frac{4}{135}-\frac{5\pi}{576},\\
&=2.35886\times10^{-3}.
\end{aligned}
\end{equation}
The longitudinal weak-saddle decrement in the same limit is $\Gamma_0^{\rm el}=G_{\rm el}/k^2=k\chi\nu/2$. Consequently,
\begin{equation*}
\frac{\delta\Gamma_\perp^{\rm el}}{\Gamma_0^{\rm el}}
\simeq
1.50170\times10^{-3}\,
\frac{\kappa}{k}\,g(v)\,\nu^3,
\qquad \nu\ll1.
\end{equation*}

\subsection{Collisionless OCP}

For the collisionless RPA spectrum in Eq.~\eqref{eq:dynamic_potential_spectrum}, let $\zeta=v/v_{\rm th}$, $p=\lambda/\kappa$, and $y=\zeta p=v\lambda/\omega_{\rm p}$. For real $\xi$ define
\begin{equation*}
\begin{aligned}
A(\xi)&=\operatorname{Re}[1+\xi Z(\xi)],\\
B(\xi)&=\operatorname{Im}[1+\xi Z(\xi)]
=\sqrt{\pi}\,\xi e^{-\xi^2}.
\end{aligned}
\end{equation*}
In the slow limit $\zeta\ll1$, Eq.~\eqref{eq:DeltaJ_dynamic_spectral} gives
\begin{equation*}
\Delta J_{\rm OCP}(\lambda)
=C\kappa^2\zeta H_{\rm OCP}(y)+O(\zeta^3),
\end{equation*}
where
\begin{equation*}
\begin{aligned}
H_{\rm OCP}(y)
={}&\frac{4}{3\sqrt{2\pi}}\int_0^\infty dP\,P^3
\Bigg\{
\frac{1}{(1+P^2)^2}\\
&-\frac{\exp[-y^2/(2P^2)]}
{\left[P^2+A\!\left(y/(\sqrt2P)\right)\right]^2
+B^2\!\left(y/(\sqrt2P)\right)}
\Bigg\}.
\end{aligned}
\end{equation*}
The two terms in this expression must be combined before the $P$ integration because their ultraviolet logarithms cancel. As an analytic check,
\begin{equation*}
H_{\rm OCP}(y)
=\frac{\pi-3}{9\sqrt{2\pi}}\,y^2+O(y^4),
\qquad y\to0,
\end{equation*}
whereas $H_{\rm OCP}(y)=[4/(3\sqrt{2\pi})]\ln y+O(1)$ for $y\to\infty$. The transverse determinant is therefore convergent at both limits.

The remaining integral is evaluated numerically. Near a weakly damped plasma pole, let $P_r$ be a root of $F_y(P)=P^2+A[y/(\sqrt2P)]$ and $\xi_r=y/(\sqrt2P_r)$. The narrow resonant contribution is resolved using the Lorentzian limit $[F_y^2+B^2]^{-1}\to\pi\delta(F_y)/|B|$, which yields $\sqrt{\pi}P_r^3/[\xi_r|F_y'(P_r)|]$ for the resonant part of the second $P$ integral before the overall prefactor. With this contribution included,
\begin{equation*}
I_\perp^{\rm OCP}
=\int_0^\infty dy\,
\left(\frac{H_{\rm OCP}(y)}{y^2}\right)^2
\simeq4.91396\times10^{-3}.
\end{equation*}
Therefore
\begin{equation}\label{eq:deltaGamma_transverse_ocp}
\delta\Gamma_\perp^{\rm OCP}
\simeq
7.82082\times10^{-4}\,
\kappa\chi^2
\left(\frac{v}{v_{\rm th}}\right)^5,
\qquad v\ll v_{\rm th}.
\end{equation}
Since $\Gamma_0^{\rm OCP}=k\chi\zeta/\sqrt{2\pi}$ on the slow weak branch, this is equivalently
\begin{equation*}
\frac{\delta\Gamma_\perp^{\rm OCP}}{\Gamma_0^{\rm OCP}}
\simeq
4.91396\times10^{-3}\,
\frac{\kappa}{k}\,g(v)\,
\left(\frac{v}{v_{\rm th}}\right)^3.
\end{equation*}

\subsection{Physical velocity dependence and range of validity}

The factors $\nu^5$ and $\zeta^5$ in Eqs.~\eqref{eq:deltaGamma_transverse_el} and \eqref{eq:deltaGamma_transverse_ocp} arise from the slow-response expansion before imposing the dispersion relation between $k$ and $v$. For a nonrelativistic particle of fixed mass whose energy is varied, $k=mv/\hbar$ and therefore $\chi=C/(\hbar m v^3)$. Eliminating $k$, the leading transverse corrections become
\begin{equation*}
\begin{aligned}
\delta\Gamma_\perp^{\rm el}
&\simeq
3.75424\times10^{-4}
\frac{C^2}{\hbar^2m^2D^5\kappa^4}\frac{1}{v},\\
\delta\Gamma_\perp^{\rm OCP}
&\simeq
7.82082\times10^{-4}
\frac{\kappa C^2}{\hbar^2m^2v_{\rm th}^5}\frac{1}{v}.
\end{aligned}
\end{equation*}
The longitudinal slow-regime decrements have the same $1/v$ dependence,
$\Gamma_0^{\rm el}=C/(2\hbar^2D\kappa v)$ and
$\Gamma_0^{\rm OCP}=C/(\sqrt{2\pi}\hbar^2v_{\rm th}v)$. Hence their ratios are velocity independent to leading order,
\begin{equation}\label{eq:transverse_dynamic_ratios}
\begin{aligned}
\frac{\delta\Gamma_\perp^{\rm el}}{\Gamma_0^{\rm el}}
&\simeq
7.50848\times10^{-4}\,
\frac{C}{m^2D^4\kappa^3},\\
\frac{\delta\Gamma_\perp^{\rm OCP}}{\Gamma_0^{\rm OCP}}
&\simeq
1.96039\times10^{-3}\,
\frac{\kappa C}{m^2v_{\rm th}^4}.
\end{aligned}
\end{equation}
The expansion of the logarithm in Eq.~\eqref{eq:deltaGamma_transverse_dynamic} is controlled in the same overlap regime: its effective small parameter scales as $\chi\nu^3=2g(v)\nu^2$ for the diffusive electrolyte and as $\chi\zeta^3=\sqrt{2\pi}\,g(v)\zeta^2$ for the collisionless OCP. It is therefore parametrically small when the slow-response and weak-saddle conditions are simultaneously satisfied. The slow-particle limit alone, however, does not generate an additional small parameter for the rectilinear approximation. Rather, the transverse correction is controlled when the slow-response condition, $g(v)\ll1$, and the spatial eikonal condition $k\gg\kappa$ possess a common domain. As discussed in Sec.~\ref{sec:electrolyte_overlap}, the kinematic part of this overlap is already absent for electron probes in ordinary liquid electrolytes, independently of the smallness of the transverse determinant. In the opposite fast limit the dynamic spectrum approaches the equal-time quasi-static result and Eq.~\eqref{eq:deltaGamma_transverse_dynamic} continuously reduces to the transverse determinant of Ref.~\cite{budkov2026static}.
\section*{Acknowledgments}
This work was supported by the Ministry of Science and Higher Education of the Russian Federation. The author thanks A.M. Satanin, L.N. Shchur, and P.E. Brandyshev for valuable discussions.


\end{document}